\documentclass[bibtex,autowc,nonblind]{apsr_submission_pseudo} 

\title{Lobbying Influence - The Role of Money, Strategies and Measurements}
\author{Fintan Oeri}{Faculty of Business and Economics, University of Basel}{\singlespacing Fintan Oeri (fintan.oeri@unibas.ch) is a PhD candidate at the Faculty of Business and Economics, University of Basel, Switzerland. Adrian Rinscheid is International Postdoctoral Fellow and Lecturer at Institute for Economy and the Environment, University of St.Gallen, Switzerland. Aya Kachi is an Associate Professor at the Faculty of Business and Economics, University of Basel, Switzerland.}
\author{Adrian Rinscheid}{Institute for Economy and the Environment, University of St.Gallen}{}
\author{Aya Kachi}{Faculty of Business and Economics, University of Basel}{}

\thanks{\singlespacing This project was part of the Swiss Competence Center for Energy Research (SCCER CREST) of the Swiss Innovation Agency, Innosuisse. We would like to thank Stefanie Bailer, Mert Duygan, Roy Gava, Denise Traber, Fr\'{e}d\'{e}ric Varone, Steven Eichenberger as well as audiences at various conferences and speaker series for their useful comments. Anne Rukkers and Julia Tran each provided excellent research assistance.}

\runningtitle{How Economic Resources, Lobbying Capacity and Group Type Shape Lobbying Influence}
\runningauthor{Fintan Oeri, Adrian Rinscheid, and Aya Kachi}

\usepackage{amsmath}
\usepackage{graphicx} 
\usepackage{rotating}
\usepackage{pdflscape}

\usepackage{tabularx} 
\usepackage{enumitem} 
\usepackage{graphicx} 
\usepackage{tabularx} 
\usepackage{ltablex} 

\usepackage{booktabs}
\usepackage{longtable}
\usepackage{array}
\usepackage{multirow}
\usepackage{wrapfig}
\usepackage{float}
\usepackage{colortbl}
\usepackage{pdflscape}
\usepackage{tabu}
\usepackage{threeparttable}
\usepackage{threeparttablex}
\usepackage[normalem]{ulem}
\usepackage{makecell}
\usepackage{xcolor}
\usepackage{enumitem}

\usepackage[utf8]{inputenc}

\usepackage{soul} 
\usepackage{lineno}
\usepackage{xfp}
\usepackage{tikz}
\usepackage{makecell}
\usepackage{url}

\usepackage[demo,   
            export]{adjustbox}
\usepackage{rotating}
\usepackage{booktabs, makecell, tabularx}

\begin{document}
\begin{frontmatter}

\bigskip \bigskip 

\begin{abstract}
Comparing the results for preference attainment, self-perceived influence and reputational influence, this paper analyzes the relationship between financial resources and lobbying influence. The empirical analysis builds on data from an original survey with 312 Swiss energy policy stakeholders combined with document data from multiple policy consultation submission processes. The results show that the distribution of influence varies substantially depending on the measure. While financial resources for political purposes predict influence across all measures, the relationship is positive only for some. An analysis of indirect effects sheds light on the potential mechanisms that translate financial resources into influence.
\end{abstract}
\end{frontmatter}

\section{Introduction}
To explain the differences in political influence, researchers often refer to the financial resources that an organization possesses \citep[e.g.,][]{Stratmann2006,Cigler2019}. Even though there is little empirical research on the causal mechanism of this relationship, the argument makes intuitive sense. Organizations with plenty of resources are freer to choose their lobbying activities and can focus on those strategies that promise the most influence. An obvious example is expensive political advertising campaigns, which are only available to wealthy political actors. Similarly, financial resources allow organizations to gather more and better information. This can open doors or even enable direct influence. 

In any case, the implementation of most political strategies should benefit from larger funds, and hardly any strategy will suffer from greater financial resources. Nevertheless, research on the nexus between financial resources and political influence is characterized by its contradictory results. Several reasons can be put forward for this.

To begin with, assessments of financial resources for political purposes, or \textit{political budget}, often rely on imprecise measurements. Primarily, this can be attributed to the fact that such information is rarely freely available. If at all, such information is shared only under assurances of confidentiality. As a result, researchers often use other measures such as staff size \citep[e.g.,][]{Kluver2011TheUnion,Yackee2020}. The total resources of an organization are another popular alternative \citep[e.g.,][]{Mahoney2007LobbyingUnion,McKay2012,Kluver2013LobbyingUnion}. 
In the case of an organization that uses all of its funds for political purposes, such an operationalization may not be problematic. However, there are many observations where this can distort the relationship between \textit{political budget} and influence. For example, a firm whose expenditures for political purposes may represent only a very small portion of its total budget. Here, the use of total funds introduces a lot of noise. It follows that under such an operationalization, it may be difficult to establish a link between financial resources for political purposes and the influence of an organization.

A similar issue arises in measuring lobbying influence. The problem starts with the fact that there is no generally accepted definition of the concept of influence. While plenty of contributions go without explicitly defining the concept, those that do tend to refer to concepts of power. Indeed, a common definition of influence is not only missing in this study's narrow context of \textit{lobbying influence} but with studies on \textit{political influence} in general. The pluralist tradition typically proceeds from Dahl's (\citeyear{Dahl1957}, pp. 202--203) simple but influential notion that ``A has power over B to the extent that he can get B to do something that B would not otherwise do". Proceeding from this conceptualization, many studies focus on control over policy outputs \citep[e.g.,][]{Dur2007,Helboe2013,Stevens2020}. The elitist response to pluralism was that a focus on open contestation neglects the less visible dimensions of power. Works in this tradition maintain that influential actors might exploit power asymmetries to prevent some issues from appearing on the political agenda (\cite{Bachrach1962}, p. 948; \cite{Schattschneider1960}, p. 69). While Schlozman (\citeyear{Schlozman1986OrganizedDemocracy}) directed the focus to the bases to exercise influence, other scholars \citep[e.g.,][]{Finger2019,Lowery2013} 
followed Lukes' (\citeyear{Lukes1974Power:Viw}, p. 28) focus on the three faces of power emphasizing also actors' ability to shape others' ``perceptions, cognitions, and preferences in such a way that they accept their role in the existing order of things".

Returning to \textit{lobbying influence} in particular, another challenge lies in the fact that influence necessarily implies a causal mechanism between intention, action and effect. Thus, in essence, the construct already requires the researcher to capture multiple variables. This is further complicated by the fact that certain actors would prefer to keep their own influence veiled. It is not surprising, then, that very different approaches to measuring influence can be found in the literature, and that the results of the respective studies vary considerably. To sum up, while the issue  has received a lot of attention in the literature, the relationship between \textit{political budget} and influence continues to be characterized by ambiguity. To a large extent, this can be attributed to different conceptualizations and operationalization of both, \textit{political budget} and influence. 

This study is set out to address these issues and provides new insights into some of these aspects. First, comparing the most common measures of lobbying influence, we shed light on how these different types of measures correspond. Whereas most previous research papers apply one or at most two types of measurements, we contrast the distribution of three different measurements of influence. Our unique data collection allows us to compare \textit{self-perceived influence}, \textit{reputational influence} and \textit{preference attainment}. The results show that the three variables are distributed quite differently. While there are signs of a common trend between the \textit{self-perceived} and \textit{reputational influence}, there is much to suggest that \textit{preference attainment} measures a rather distinct construct. 

Second, we compare the explanatory power of \textit{political budget} with respect to the different types of measurement. In this context, it is important to note that our analysis explicitly focuses on financial resources for political purposes. However, our data allow for a comparison with total resources and the number of employees commissioned to follow political events, two popular alternative operationalizations. To accommodate the different scales of measurement of influence, we run separate regressions for \textit{self-perceived influence}, \textit{reputational influence} and \textit{preference attainment}. Again, the empirical analysis points to substantial differences between the types of measurements. \textit{Political budget} shows a statistically significant association for all three measures. Indeed, the relationship with \textit{political budget} appears to be very similar for the two perceptional measures, i.e., \textit{self-perceived influence} and \textit{reputational influence}. However, while significant, the relationship with \textit{political budget} is negative for \textit{preference attainment}.

Third, we provide new evidence on the causal mechanism that allows \textit{political budget} to have an impact on lobbying influence. We examine to what extent lobbying activities translate the effect of \textit{political budget} on influence. Following the approach of \cite{Imai2011UnpackingStudies}, we apply a counterfactual framework to identify these indirect effect through mediation analysis. The findings show that the indirect effects of \textit{political budget} on lobbying influence run through specific activities. In line with the other findings, the results suggest that there are important differences between the three measures of influences.

The following section introduces the relevant debates in the lobbying literature, maps out our research questions and introduces our core contributions. This is followed by the third part, which presents the research design including the data gathering process, variable operationalization and an introduction to the methods applied. The following section presents the findings from the empirical analyses while the subsequent discussion provides some interpretation and connects the results back to the relevant scientific discussions. Finally, the paper concludes with a summary of the main implications of our work and provides an overview of additional aspects that deserve attention in future research.


\section{Background}
\subsection{Literature review}

The subsequent analysis relates to different, yet related literatures. The ultimate goal is to reveal the causal mechanism of the effect of \textit{political budget} on an organization's lobbying influence. However, the analysis of this causal mechanism requires a critical assessment of the relationship between \textit{political budget} and lobbying influence. For this, in turn, it is central to consider the scientific evidence related to the measurement of lobbying influence.

\subsubsection{Measurement of influence}
Accordingly, we begin by addressing the state of research on the measurement of lobbying influence. The literature distinguishes between three different types of quantitative lobbying influence measures: \textit{self-perceived influence}, peer-attributed or \textit{reputational influence}, and \textit{preference attainment}.

So-called \textit{self-perceived influence}, i.e., measuring lobbying influence by asking stakeholders to evaluate their own influence, finds widespread use in the literature \citep[e.g.,][]{Helboe2013,Lyons2020,McKay2012,Newmark2017,Stevens2020,Yackee2020}. Capturing \textit{reputational influence} is more labor-intensive and less common \citep[e.g.,][]{Finger2019,Stevens2020}. Typically, peers or experts assess the lobbying influence of stakeholders. The two methods share two important characteristics. First, they are both characterized by the fact that the measured quantity is not lobbying influence as such, but perceptions thereof. Second, both, \textit{self-perceived} and \textit{reputational influence}, presuppose a causal mechanism between intention and effect.   

Both properties contrast with the third measure, \textit{preference attainment}, which has come to be widely used too \citep[e.g.,][]{Mahoney2007LobbyingUnion,Dur2008,Nelson2012LobbyingRulemaking,Kluver2013LobbyingUnion,Bernhagen2014}. Here, the size of the overlap between individual preferences and enacted policies serves as a measure to distinguish the winners from the losers of a policy process. What is missing, however, is the requirement of a causal relationship. An actor can see its preferences fulfilled without having contributed anything to such an outcome. Rather than calling it lobbying influence, some scholars therefore refer to it as lobbying success \citep[e.g.,][]{Bernhagen2014}.

It follows that none of the three variables measure lobbying influence directly. Table \ref{t_measuresoverview} offers an overview of the advantages and disadvantages of the three measures.%

\begin{table}[!h]\centering\footnotesize
\renewcommand{\arraystretch}{1.2}
\begin{tabular}{ @{}p{0.16\textwidth} p{0.25\textwidth} p{0.25\textwidth} p{0.25\textwidth}@{}}
& Self-perceived influence
& Reputational influence
& Prefence attainment \\
\midrule
Underlying concept
& Perception of influence over policy
& Perception of influence over policy
& Policy goal attainment \\[.5cm]
Advantages
& \begin{itemize}[topsep=0pt, partopsep=0pt, leftmargin=*]
      \item Inherent causal link between intention (or actions) and outcome
      \item Simple data gathering
      \item Covers insights from all channels of influence exertion
\end{itemize}
& \begin{itemize}[topsep=0pt, partopsep=0pt, leftmargin=*]
    \item Inherent causal link between intention (or action) and outcome
    \item Same data source for every value
    \item Risk from one biased perception mitigated
\end{itemize}
& \begin{itemize}[topsep=0pt, partopsep=0pt, leftmargin=*]
    \item Objective assessment
    \item No discrimination between different channels of influence
\end{itemize} \\
Challenges
& \begin{itemize}[topsep=0pt, partopsep=0pt, leftmargin=*]
    \item Perception as a subjective assessment
    \item Risk of strategic responses
    \item Different data source for every value (comparability of scale)
\end{itemize}
& \begin{itemize}[topsep=0pt, partopsep=0pt, leftmargin=*]
    \item Perceptions as a subjective assessment
    \item High costs of data gathering
    \item Risk of overweighing channels of influence that are easily observable
\end{itemize}
& \begin{itemize}[topsep=0pt, partopsep=0pt, leftmargin=*]
    \item Lack of causal link between intention (or actions) and outcomes
    \item Requires truthful identification of preferences
\end{itemize} \\
References
& e.g., \cite{McKay2012,Helboe2013,Newmark2017}
& e.g., \cite{Finger2019,Stevens2020}
& e.g., \cite{Dur2008,Bunea2013IssuesPolicy,Gilens2014TestingCitizens,Bernhagen2014} \\
\bottomrule
\end{tabular} %
\caption{Measures of lobbying influence - overview of conceptual and empirical properties.} 
\label{t_measuresoverview}
\end{table}

One strength of \textit{reputational influence} is that it is typically based on a number of data sources, for example peers or a group of experts, and thus individual, biased judgments carry less weight. However, this type of data gathering is also associated with high costs. Moreover, the measurement runs the risk of overestimating types of influence that are particularly visible. Actors that engage in lobbying activities that create public attention, e.g., major advertising campaigns, may be considered more influential than organizations that take a less prominent role in the public perception but may be equally effective, e.g., by lobbying decision makers directly. 

In the case of \textit{self-perceived influence}, the relative simplicity of the data gathering is an important advantage. Also, in contrast to \textit{reputational influence}, the assessment builds on first-hand information: regarding their own activities and immediate consequences thereof, actors have outstanding knowledge. However, there is a risk that actors may deliberately provide inaccurate or strategic responses concerning their own influence. In an attempt to give a greater picture of their relevance and legitimacy, organizations may be tempted to provide inflated accounts of their influence \citep[e.g.,][]{Lyons2020}. Other organizations may feel tempted to underestimate their influence to avoid negative reactions \citep[e.g.,][]{Stevens2020}.

A similar challenge arises with \textit{preference attainment}, namely the truthful representation of preferences \citep[e.g.,][]{Dur2008}. Such data can be obtained by asking experts or stakeholders to indicate other stakeholders' policy preferences or 'ideal points' \citep[e.g.,][]{Bernhagen2014,DeBruycker2019}, asking stakeholders to indicate their own policy preferences \citep[e.g.,][]{Mahoney2007LobbyingUnion,Baumgartner2009a}, or determining preferences from policy documents \citep[e.g.,][]{Yackee2006ABureaucracy,Kluver2011TheUnion}. In the context of the two latter methods, there may be an incentive to misrepresent true preferences leading to distorted estimates if an actor anticipates later negotiating concessions \citep[e.g.,][]{Dur2007,Lowery2013}. 

The lobbying influence of other involved parties represents another issue. An actor, whose lobbying efforts made no difference, may nonetheless be considered successful because for some unrelated reasons, e.g., lobbying of more influential actors, the actor attained all her policy goals. At the same time, an actor may have been influential without achieving policy goals, e.g., when the actor's influence prevented an even less desirable outcome\footnote{\cite{Helboe2013} addresses this problem by linking influence more closely to specific activities (e.g., questions to parliamentary committees) instead of preference statements, and attaches greater relevance to the chronological sequence. However, this thwarts one of the strengths of \textit{preference attainment}, namely that success can be measured regardless of the channel of influence \citep{Dur2008}. In addition, such an approach requires considerable effort in the coding of activities, which may prevent its large-scale application in the context of \textit{preference attainment}.}.

An important advantage of the preference attainment method, however, is that it can be based on publicly available data such as consultation submissions and legislative texts. In addition, some scholars emphasize the fact that it is not a matter of perceptions. Rather, the measure builds on the factual congruence between stated preferences and the policy outcome. This also means that the method is insensitive to the channels of influence. It does not matter whether influencing takes place in a back room or in the course of a large-scale advertising campaign; only the visible results in terms of policy outcomes count.


\subsubsection{Political budget and influence}

With the differences between influence measures in mind, we move to another set of studies, which address the explanatory power of potential predictor variables. Here, the question arises as to what extent potential differences between the measures may also be reflected in the corresponding explanatory approaches. The role of \textit{political budget} on the one hand, and \textit{business interests} on the other hand are of particular interest.

Two things stand out when looking at previous contributions on the effect of resources on lobbying influence. First, there are substantial differences in the operationalization of \textit{political budget}. With the sum of reported lobbying expenditures, \cite{Junk2019WhenCoalitions} not only employs the variable that is most similar to ours. Arguably, it is also the most precise measurement. Other operationalizations can be expected to introduce a higher degree of noise into the relationship between an actor's resource endowment and lobbying influence. For example, \cite{McKay2012} uses total resources (revenue, budget or sales) while a number of studies use staff size. However, there are differences here as well, as this can be based on the total number of employees \citep[e.g.,][]{Kluver2013LobbyingUnion} or the employees entrusted with lobbying tasks \citep[e.g.,][]{Mahoney2007LobbyingUnion,Stevens2020}.  

Second, and possibly resulting from these differences in the operationalization, the results on the explanatory power of financial resources are mixed. For example, 
\cite{Junk2019WhenCoalitions} finds no noteworthy effect on \textit{preference attainment}. This is in line with multiple other studies using other operationalizations of financial resources \citep[e.g.,][]{Mahoney2007LobbyingUnion,Baumgartner2009a}. \cite{Kluver2013LobbyingUnion}, however, goes beyond the individual organization as the unit of observations and analyzes the relevance of lobbying camps. Both, the aggregate economic power at the lobbying camp level and the individual level, exhibit statistically significant associations with \textit{preference attainment}. Interestingly, however, the association is positive only in the first case, while the probability of attaining preferences decreases with individual economic power. With respect to measures other than preference attainment, \cite{McKay2012} finds organizations' economic resources to explain hardly any variance neither in \textit{preference attainment} nor \textit{self-perceived success} (not influence). \cite{Stevens2020}, on the other hand, use a composite measure building on \textit{self-perceived} and \textit{reputational influence} and find a positive association with economic resources.

These contradictory findings may result from the use of different measurement methods. However, even among studies using the same measurement concepts, there seems to be substantial disagreement. In any case, these differences regarding the measurement of influence as well as \textit{political budget} make it difficult to compare findings across studies. Further complicating matters, the studies also differ with regard to other aspects like country, policy venue or policy content.

Turning to the other major predictor, whether an organization represents \textit{business interests}, it is important to note that the literature often associates a larger \textit{political budget} with this type of interest \citep[e.g.,][]{McKay2012,Varone2020}. In fact, some contributions go as far using the business/non-business dichotomy as a proxy for resources \citep[e.g.,][]{Binderkrantz2014AConsultations}. However, resources have repeatedly been found not to vary systematically with group type \citep[e.g.,][]{Baroni2014DefiningGroups,Kluver2012BiasingPolicy-Making}. Irrespective of the relationship between group type and resources, multiple contributions suggest an elevated responsiveness of policymakers to requests from groups representing \textit{business interests} \citep[e.g.,][]{Gilens2014TestingCitizens,Rasmussen2014DeterminantsConsultations,Balles2020SpecialAttention,Varone2020}. In any case, \textit{business interests} represent an important predictor that is taken into account in most empirical analyses of influence \citep[e.g.,][]{Yackee2006ABureaucracy,Dur2019TheUnion}.



\subsubsection{The causal mechanism}

Finally, the third avenue of research relevant for this study focuses on the mechanism through which \textit{political budget} may affect lobbying influence. Generally, in line with how rational choice scholars conceptualize lobbying, the argument rests upon the assumption that resources can be converted to something that is of value to decision makers \citep{Stigler1971TheRegulation}. For example, many political decisions build on insights from actors affected \citep{Austen-Smith2009}. Developing the capacity to prepare and provide such expert information usually is easier with higher financial means \citep[e.g.,][]{Hall2006LobbyingSubsidy}.
While there may be exceptions, typically, stakeholder cannot affect policy outcomes simply by possessing a lot of money. Resources need to be mobilized to unfold their potential impact in terms of influencing policy \citep[e.g.,][]{Duygan2021IntroducingRegimes}.

In this context, lobbying strategies describe the activities that interest groups engage in to influence policies according to their preferences. While this also refers to the decision whether to join forces or lobby alone \citep[e.g.,][]{Nelson2012LobbyingRulemaking}, the literature often focuses on insider and outsider strategies \citep[e.g.,][]{Mahoney2007LobbyingUnion,DeBruycker2019}. Surprisingly little research has been conducted on how these activities mediate the effect of resources on lobbying influence, with \cite{Binderkrantz2019TheUK} representing a notable exception. Using group type as a proxy for different resource endowments, they distinguish between insider and outsider strategies. The empirical analyses reveal that in addition to a direct effect of group type on influence, there indeed is also an indirect effect running through this choice of lobbying strategy.

In another contribution to this topic, \cite{McKay2012} finds that a range of lobbying activities predict lobbying success but only a subset of them are associated with higher financial means, namely engaging in multiple venues as well as contributing and participating in PACs. It is worth noting that the analysis of these indirect effects builds on a rather weak association between resources and lobbying success. Also, while mediation analysis has experienced substantial development since, the empirical approach in \cite{McKay2012} builds on pairs of bivariate relationships only. 


\subsection{Research questions}

Based on these gaps in the literature, this study addresses the following three research questions. First, comparing \textit{self-perceived influence}, \textit{reputational influence} and \textit{preference attainment}, we conduct an empirical investigation to test whether (and to what extent) the patterns of influence emerging from the three measures overlap. Are there substantial differences between the three measures, or do actors, which are influential according to one measure, also score high on the two other measures?

Second, to what extent can an organization's \textit{political budget} explain its influence scores? Building on the insights from the first analysis on the comparability of measures, we analyze whether the explanatory power of \textit{political budget} corresponds between influence measures.

Third, what lobbying activities, if any, transmit a potential effect of \textit{political budget} on influence? Again, we compare the findings to assess the congruence between the three measures.



\subsection{Contributions}

Our contribution has broader implications for our understanding of how interest groups influence public decision-making. This fundamental topic has long been of interest to many political scientists and economists across subfields  \citep[e.g.,][]{Salisbury1989WhoLobbyists,Wawro2001,Hall2006LobbyingSubsidy,Box-Steffensmeier2013QualityMakingb,Gilens2014TestingCitizens,Schnakenberg2016,Box-Steffensmeier2019Cue-TakingLetters,Bertrand2014IsProcess,Bertrand2020,Schnakenberg2021}. 
Over the past decades, the topic has gained even more practical relevance as debates around urgent policy agendas have become increasingly contested and politicized \citep{Lupia2013a,Bouleau2019PoliticizationEnvironment,Fowler2015,Druckman2017UsingEffective}. 
As such, scholars have emphasized the importance of investigating the politics---especially the working of lobbying---behind policymaking, for instance, in migration \citep{Spirig2021}, 
climate change \citep{stokes2016electoral,goldberg2020OilEnvironment,Farrell2016CorporateChange,kim2014electric,Cory2021SupplyCoalition}, energy system transitions \citep{Stutzer2021,Hughes2020,Duygan2021IntroducingRegimes}, and public health \citep{Bowers2018HowImplants,Wouters2020,Harris2021}. 
Inspecting what it means for competing stakeholders to be politically influential and how the distribution of influence is linked to their endowments as well as lobbying practices thus leads to a better understanding of the nature of political contestations. Resulting insights may help to explain why democratic policy processes often lead to a stalemate in critical policy areas.   

Much of the prior literature probing the resource-influence nexus has been muddled by conflicting empirical findings due to both data limitations and fragmented research targets.
In the first place, actors' \emph{influence} is notoriously hard to operationalize. 
With an explicit focus on the differences between measures, this study goes beyond previous contributions by comparing three measures of lobbying influence. %
Moreover, good data are hard to obtain also for the financial resources allocated to political purposes. As a result, each existing study has resorted to various proxy measures. 
In contrast, the uniform setting of our study allows for a reasonably standardized comparison using a precise measure of \emph{political budget} as well as alternative operationalizations across the three different measures of influence. %

\begin{figure}[hbt!]
\includegraphics[width=.96\textwidth]{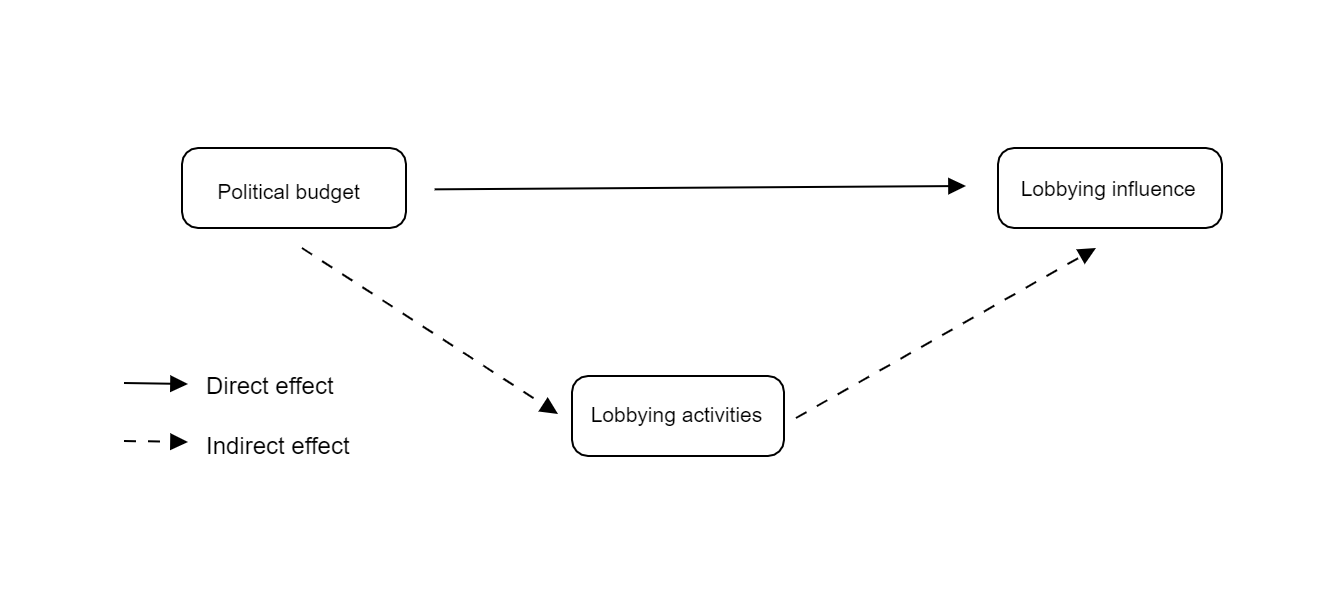}
\caption{Mediation framework.}
\label{f_framework}
\end{figure}

Concerning the target of empirical analysis, at least since Berry's groundbreaking study of American public interest groups \citep{Berry1977}, prior literature has often conjectured on the role of interest groups' lobbying \emph{practices} (e.g., inside and outside lobbying strategies) 
as relevant pieces of the puzzle. However, few studies have explicitly embedded these factors in their analyses. 
Consequently, in our view, the literature overall seems to suggest---and it is indeed reasonable to theorize---that lobbying strategies potentially mediate the effect of \emph{political budget}  on the actors' political influence 
and yet, existing data and approaches do not allow us to test these relationships in detail. We argue that typically, the effect of resources on policy influence is mediated by the activities the actor engages in. In other words, wealth needs to be put to use, e.g., by gathering information and providing it to decision makers. Figure \ref{f_framework} illustrates the relationships considered in this study.

\begin{figure}[hbt!]
\includegraphics[width=.96\textwidth]{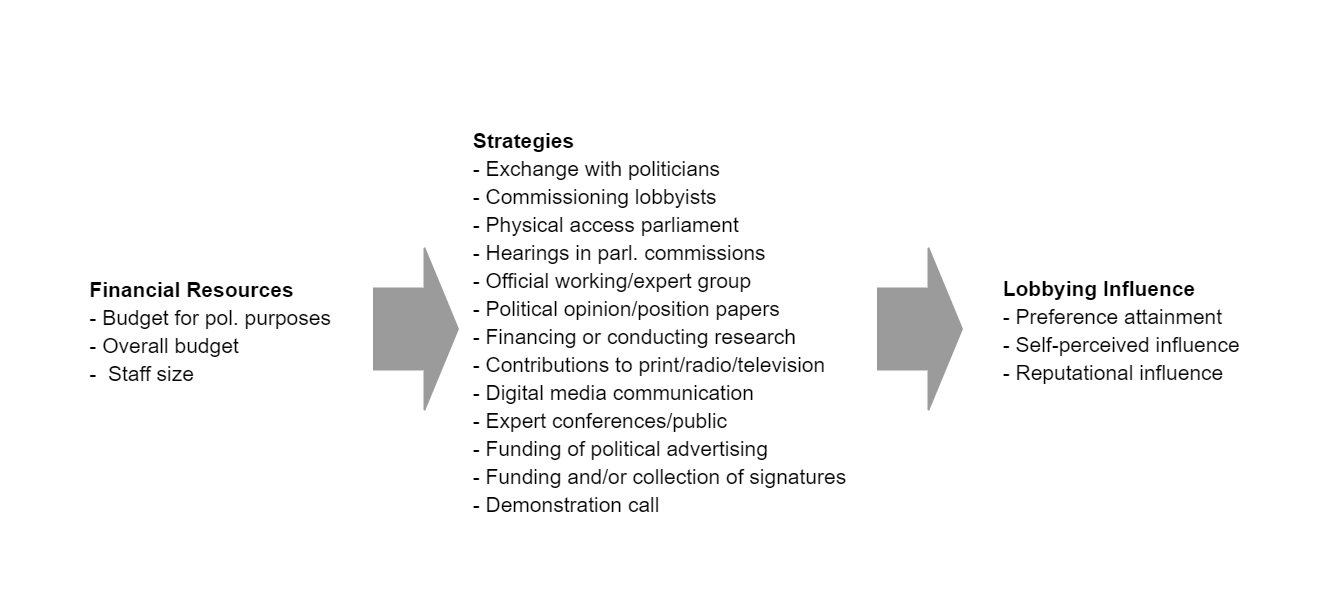}
\caption{Indirect effects.}
\label{f_indirect}
\end{figure}

Our study benefits from a novel dataset with a comprehensive set of three influence measures (each corresponding with the concept employed in previous studies), a variable precisely measuring the \emph{political budget}, and an empirical framework that allows stakeholder activities (each of which belongs either to an inside or outside lobbying strategy) to mediate the effect of resources on influence. 
Thus, not only is our study the first to provide a direct comparison of the three influence measures most often used in the lobbying literature. It also assesses whether and to what extent \emph{political budget} is linked to the actors' influence and analyzes the relevance of the lobbying practices for this relationship. 

\newpage

\section{Empirical strategy}
\subsection{Analysis of political influence}\label{methods}
In this section, we lay out the empirical strategy to address the research questions. This is followed by a short introduction to our case. Finally, we describe our data collection strategy and explain how we operationalized the constructs outlined above.

To address the question about the effect of \textit{political budget} on lobbying influence, we run three separate regressions; one for each measure of influence. Because of the differences in measurement scales (see subsection on \nameref{data_operationalisation}), we use three different regression models. 

For \textit{self-perceived influence}, we apply a standard linear regression model. As in both other models, the main interest lies in organization-specific \textit{political budget}. As a control, we include a variable indicating whether the actor represents \emph{business interests}.

In the case of \textit{reputational influence}, we estimate a zero-inflated count data regression model, which returns a zero- and a count model \citep{Zeileis2008}. We used the same set of predictors in both models. The zero-model, estimating the probability of being mentioned by no survey participant, uses logistic regression. To explain the variance among non-zero outcomes, the count model, then, uses Poisson regression. 

In contrast to the perceptional measures, the observation unit for the \textit{preference attainment} measure is on the subject-item level. Hence, in addition to the predictors used in the other specifications, we included two random intercepts to account for potentially correlated errors among observations for the same organizations as well as for the same items of a bill. We estimate the model using mixed effects logistic regression. 

While the regression results may explain whether an organization’s \textit{political budget} affects its influence over energy policy, it provides little insight on how such a potential effect comes about. On the condition that the regression results identify such an effect, in the final step of our analysis, we shed light on the causal mechanism that leads \textit{political budget} to affect policy influence.

To that end, we estimate to what extent the effect of \textit{political budget} is mediated by various lobbying activities. The analysis follows the approach of \cite{Imai2010IdentificationEffects}. It applies a potential outcomes framework and builds on two separate models. First, we estimate the mediator model featuring the mediated variable, i.e., \emph{political budget}, as well as potential confounders, i.e., lobbying activities, as predictors. Second, along with the same set of independent variables, the predicted values of the mediator, then, serve to fit the model of the actual outcome variable, which in our case is influence. From this second model, the average difference in the predicted outcomes for different values of the mediator, holding constant the value of the mediated variable, represents the average causal mediation effect (ACME). 

We conduct this mediation analysis for each potential causal mechanism separately using the \textit{mediation} and the \textit{maczic} package \citep{Tingley2014Mediation:Analysis,Cheng2018MediationData}. Under the assumption that the lobbying activities are not causally related, the ACMEs can be consistently estimated without explicitly including other mediators \citep{Imai2013IdentificationExperiments}. This does not hold for the average direct effect (ADE), which tends to be overestimated if not all indirect effects are being considered simultaneously. However, the goal of the analysis is to determine the mediated effect. 
Finally, we estimate the confidence intervals using the default quasi-Bayesian Monte Carlo method, as this was the only option available for all three types of regressions. However, the results were robust to the use of non-parametric bootstraps for the perceptional measures, where this estimation of the confidence intervals was possible (see Table \ref{app_t_mediation.boot} and Figure \ref{app_f_coefplot.boot} in appendix \ref{app_bootstrap}).

An important advantage of this form of mediation analysis is that there are no requirements regarding the functional form\footnote{While this is true in theory, the technical implementation required minor adjustments in the case of the mixed model for the \textit{preference attainment} measure. More specifically, the outcome model features a random intercept only for organizations. This stands in contrast to the specification used in the main regression, which includes an additional random intercept for items. To ensure that this does not fundamentally alter the relationship between \textit{political budget} and the \textit{preference attainment} variable, model 1 of  \ref{app_t_reg.compare.pa} in appendix \ref{app_regression} features the regression results for this modified specification. While the estimates slightly vary, the directions remain the same and both predictor variables continue to enjoy statistical significance.}.
This is in contrast to the product and difference methods, which are commonly used to estimate indirect effects in linear structural equation models \citep{Baron1986TheConsiderations}. In the case of logistic regression, as repeatedly used in this study, both of these methods would fail to estimate mediation effects consistently \citep{Imai2010AAnalysis}. Moreover, the underlying identification strategies omit necessary assumptions required to establish causal mechanisms \citep{Glynn2012TheEffects}. 

Indeed, since our approach relies on a counterfactual outcome that is never actually observed, the identification of ACMEs requires the strong assumption of sequential ignorability. Its first part, the exogeneity of the mediated variable, may be considered standard. The second aspect is more restrictive, particularly in non-experimental settings like this study. Conditioning on the full set of relevant confounders and, importantly, the mediated variable, the observed mediator value is assumed to be ignorable, implying identifiability of the ACME \citep{Imai2011UnpackingStudies}. In the context of this study, this means that all variables confounding the relationship between the lobbying activity and the influence measures must be conditioned on. This includes \textit{political budget}, reflecting the sequentiality element that sets apart the causal mediation effect (sequential ignorability assumption) from a causal effect of the mediator (conventional exogeneity assumptions).

Finally, the assumption also prohibits any influence of the mediated variable, \textit{political budget} in our case, on the other confounders, no matter whether they are observed or not. This also implies that there must not be any causal relationship between different mediators, as this would also represent post-treatment confounding \citep{Imai2013IdentificationExperiments}.

Given the assumption is satisfied, ACME is identifiable even though not all potential outcomes can be observed. Unfortunately, the assumption is not testable. However, at least for some violations, sensitivity analyses can indicate to what extent the assumption needs to be violated for the mediated effects to lose statistical significance \citep{Imai2010AAnalysis}.

\subsection{Data collection}\label{data_operationalisation}

To examine the research questions introduced above, we collected data on policy actors' resources, lobbying activities, preferences, and lobbying influence. Our study is embedded in the context of Swiss energy policy, which -- as we will show below -- is uniquely suited for such an empirical investigation. Switzerland has recently experienced important energy policy processes with distinct outcomes. This includes a federal strategy to realign its national energy policy. After several years in parliament, the so-called 'Energiestrategie 2050’ was accepted by the electorate in 2017. Shortly after, a major revision of the regulation on carbon emissions (CO2 law as part of ‘Klimapolitik nach 2020’) failed to gain a legislative majority in its first attempt. The parliament then passed a revised version, which eventually was rejected in a public referendum in 2021.

Switzerland's political system is well suited for our study because the legislative process builds on the results of a highly formalized and transparent consultation submission process. Stakeholders can, and frequently do, submit their views on public policy proposals. Due to this institution, Swiss policymaking is characterized by a more ``direct incorporation of a wide variety of parties and organizations" (\citeauthor{Weiler2015}, \citeyear{Weiler2015}, p.~750) than other political systems. Importantly, given that the consultation process is open to all stakeholders and implemented transparently (e.g., online publication of stakeholders' submissions), there is a certain level playing field in terms of stakeholders' \textit{access} to the political decision-making process. This makes it all the more interesting to study whether certain political resources or strategies make some stakeholders more \textit{influential} than others.

Our empirical analysis builds on two separate data sources. To determine the \textit{preference attainment} measures, we use publicly available consultation submissions. The remaining variables, including the perceptional measures of influence, stem from a survey we conducted with Swiss energy stakeholders in 2019.

The \textit{preference attainment} values build on subject-item data from the preferences indicated in consultation submissions on the two policies mentioned above, 'Energiestrategie 2050’ and ‘Klimapolitik nach 2020’. These bills are suited for the purposes of this study for the following reasons. First, they were subject to intensive political and public debate and were both ultimately put to an optional referendum, which in both cases was filed by the Swiss people's party (SVP). Not only did this spotlight mobilize a larger set of actors. Also, the attention required public position taking by organizations that under different circumstances may prefer to hide their preferences and lobby decision makers more covertly. Similarly, the contested nature of these bills may be more informative with regard to which organizations truly are influential. To attain one side's preferences, decision makers had to let down other major stakeholders. 

Second, the two policies cover the full spectrum of energy policy topics, such as nuclear phase-out schedule, car emission limits, subsidies for renewable energies, transparency rules for power supply firms, ratification of the Paris Agreement, emission reduction targets, emission trading scheme, and fossil heating bans. With regard to the comparison of influence measures, this feature of our sample is important. A narrower set of issue areas may not be as representative to measure actor-specific influence on energy policy.

Finally, the multiple choice format of these two consultations makes it possible to collect data on preferences for a large sample in a standardized manner. Smaller consultations  usually ask participants to share their views in a more general text form. This format entails a lot more ambiguity and requires the coder to interpret the answer, leading to difficulties in extracting preferences and comparing policy positions. The lack of clearly spelled out questions further hampers comparability. 

To determine the \textit{preference attainment}, we first registered all item-specific choices for each participant. Next, we matched these preferences with the corresponding outcome from the final version of the bill that the parliament agreed upon. This resulted in a binary measure indicating whether the parliament's decisions overlapped with the stakeholder's preferred outcome. We determined this value for all item-organization observations. In case the parliamentary outcome was unclear, we dropped the item for all organizations. If an organization did not provide an answer to a specific question or answered ambiguously, we only dropped the corresponding subject-item observation. 

This approach resulted in 44 items that were considered for the \textit{preference attainment} measure\footnote{The full list of items contains 26 items from the 'Energiestrategie 2050' consultation and 18 from the ‘Klimapolitik nach 2020’ consultation. For a full list of items, see Table \ref{app_t_consultation.item.es2050} and Table \ref{app_t_consultation.item.klima} in appendix \ref{app_consultation}.}. The survey data builds on a larger sample size. In addition to the participants from the two policies mentioned above, we also considered the submissions from another two major Swiss energy policy consultations (ordinance level implementation of Energiestrategie 2050 \&  Stromversorgungsgesetz, which regulates energy supply). Taking into account that some organizations participated in more than one consultation, 740 unique participants were identified. While we excluded individuals, there were no further formal or informal conditions to qualify as a stakeholder. However, 60 organizations had to be dropped because they had ceased to exist or because we were not able to obtain neither online nor postal contact details.

To inform them about the project and the process, potential survey participants received an invitation letter with personalized details whenever possible. Based on the insights from a trial with 38 organizations, the survey underwent minor adjustments to reduce drop out rates. 

We ran the survey on the Qualtrics survey platform. Given that Switzerland consists of several language regions, respondents could choose between a German, a French, or an English version of the survey. Organizations that had not submitted a response within the two-week deadline received a reminder letter (main wave) or phone call (trial) encouraging them to participate. Enclosed with the reminder letter was a paper version of the survey and a pre-franked envelope. Respondents could either use the online link or return the hard copy within two weeks.

Overall, data collection took 20 weeks.\footnote{Given its nonrecurring nature, this type of data gathering process is susceptible to biases from topical contexts. While the climate strike movement did gain traction over this period, the authors are not aware of individual energy policy related events (e.g., referenda on energy bills), which may suggest distortions from temporal anomalies.} Out of 680 organizations invited to participate, 364 submitted a response, resulting in a response rate of 53.5\%. After removing insufficiently completed submissions, the remaining data contains 312 observations (45.8\%) on a total of 42 survey items.  

To determine \textit{self-perceived influence}, we asked respondents to assess the influence of their organization on Swiss energy policy. Respondents had to choose from an ordered six point scale ranging from ``No influence at all" to ``Very strong influence". Moreover, as a measure of \textit{reputational influence}, respondents were asked to list the organizations that they considered to be influential in Swiss energy policy. For each unique answer provided, we added up the mentions across all participants, not counting self-mentions.

As mentioned above, the remaining variables also build on survey responses. \texttt{Political budget} was measured by an organization's annual budget devoted to political purposes. In the survey, we asked respondents to choose out of seven answer options ranging from less than CHF 5'000 to more than CHF 1'000'000. The survey did not force participants to respond to this item. However, when trying to skip the item, respondents were reminded of the importance of this information. Out of the 312 organization in our sample, 269 (86\%) provided the information. Regarding the entire sample (680 organizations), this corresponds to approximately 40\%. The binary variable measuring whether an organization represented \textit{Business Interests} was hand-coded. The measure distinguishes \textit{Energy Businesses}, \textit{Other Businesses} and \textit{Business Associations} from the remaining actors.

Finally, we asked respondents about the lobbying activities their organizations had pursued in the past ten years. We provided a list of 13 activities with an additional text entry option and respondents were free to tick as many as applied to them, resulting in one binary subitem for every activity (see Table \ref{t_activities} for an overview of lobbying activities presented).  

\begin{table}[!h]\centering\footnotesize
\renewcommand{\arraystretch}{1.2}
\begin{tabular}{c p{0.9\textwidth}}
\multicolumn{2}{l}{Has your organization pursued the following activities~during the last ten years? (Multiple answers possible)} \\
\midrule
a. & Informal exchange with politicians. \\ b. & Mandating other organizations or experts to follow the political events or actively represent interests of the organisation. \\ c. & Accessing non-public parts of the national parliament building (Wandelhalle). \\ d. & Participation in hearings of parliamentary commissions. \\ e. & Participation in an official working/expert group to draft new legislation. \\ f. & Preparation and publication of political opinions and position papers. \\ g. & Financing or conducting research. \\ h. & Active involvement in media debates (eg opinion articles or interviews in print media, radio, television, etc.). \\ i. & Communication with the public via digital media (eg Facebook, Instagram, Twitter, etc.). \\ j. & Organization of expert conferences and/or public debates. \\ k. & Funding of political advertising. \\ l. & Funding and/or collection of signatures. \\ m. & Demonstration call. \\ n. & Other: \_\\
\bottomrule\\
\end{tabular}
\caption{Survey items on lobbying activities.} 
\label{t_activities}
\end{table}

\newpage
\section{Results}
\subsection{Descriptive results}
Table \ref{t_descriptives} presents descriptive statistics for all variables. One important insight is that the reduction in sample size results from missing data on \texttt{political budget}, \textit{self-perceived influence} and of course the smaller sample for the \textit{preference attainment} measure. Furthermore, the table reveals noteworthy patterns in the dependent variables. Most strikingly, in the majority of subject-item observations, preferences overlapped with the policy outcome. Although not directly comparable, this contrasts with the fact that more than half of all organizations were not named by any respondent as influential, highlighting the zero-inflated nature of the \textit{reputational influence} measure. Finally, for the \textit{self-perceived influence} measure, the mean is also slightly lower than the median suggesting that survey participants were more reluctant to label themselves as very influential compared to not influential at all. 

\begin{table}[!h]\centering\footnotesize
\renewcommand{\arraystretch}{1.2}
\begin{tabular}{@{\extracolsep{5pt}}lllrrrrr}
 & & & N & Mean & Median & Min & Max\\
\midrule
\multicolumn{3}{l}{\small \textsc{Predictors of influence}}\\
& \multicolumn{2}{l}{Business interests} & 312 & 0.58 & 1 & 0 & 1\\
& \multicolumn{2}{l}{Political budget} & 269 & 2.73 & 2 & 1 & 7\\[.25cm]
\multicolumn{3}{l}{\small \textsc{Lobbying activities (mediators)}}\\
&\multicolumn{2}{l}{Exchange with politicians} & 312 & 0.79 & 1 & 0 & 1\\
&\multicolumn{2}{l}{Commissioning lobbyists} & 312 & 0.34 & 0 & 0 & 1\\
&\multicolumn{2}{l}{Physical access parliament} & 312 & 0.28 & 0 & 0 & 1\\
&\multicolumn{2}{l}{Hearings in parl. commissions} & 312 & 0.30 & 0 & 0 & 1\\
&\multicolumn{2}{l}{Official working/expert group} & 312 & 0.41 & 0 & 0 & 1\\
&\multicolumn{2}{l}{Political opinion/position papers} & 312 & 0.60 & 1 & 0 & 1\\
&\multicolumn{2}{l}{Financing or conducting research} & 312 & 0.29 & 0 & 0 & 1\\
&\multicolumn{2}{l}{Contributions to print/radio/television} & 312 & 0.54 & 1 & 0 & 1\\
&\multicolumn{2}{l}{Digital media communication} & 312 & 0.38 & 0 & 0 & 1\\
&\multicolumn{2}{l}{Expert conferences/public debates} & 312 & 0.37 & 0 & 0 & 1\\
&\multicolumn{2}{l}{Funding of political advertising} & 312 & 0.14 & 0 & 0 & 1\\
&\multicolumn{2}{l}{Funding and/or collection of signatures} & 312 & 0.15 & 0 & 0 & 1\\
&\multicolumn{2}{l}{Demonstration call} & 312 & 0.11 & 0 & 0 & 1\\[.25cm]
\multicolumn{3}{l}{\small \textsc{Measures of Influence}}\\
&\multicolumn{2}{l}{Self-perceived influence} & 298 & 2.78 & 3 & 1 & 6\\
& \multicolumn{2}{l}{Reputational influence} & 312 & 3.20 & 0 & 0 & 118\\
& \multicolumn{2}{l}{Preference attainment} & 4957 & 0.61 & 1 & 0 & 1\\
& & \textit{Items answered} & 216 & 22.95 & 22 & 1 & 41\\
\bottomrule
\end{tabular}
\caption{Descriptive statistics.} 
\label{t_descriptives} 
\end{table}

For informational purposes, the table also includes an entry for the number of \textit{preference attainment} items per organization, which was not a variable used in any of the specification as such \footnote{Rather than presenting numbers on the whole population of consultation participants, this entry refers only to the subsample that also submitted a survey response.}. The number of observations ranges from 1 to 41 items per organizations, underlining the importance of a mixed model approach rather than using a ratio variable of preferences attained.

\begin{figure}[ht!]
\includegraphics[width=.77\textwidth]{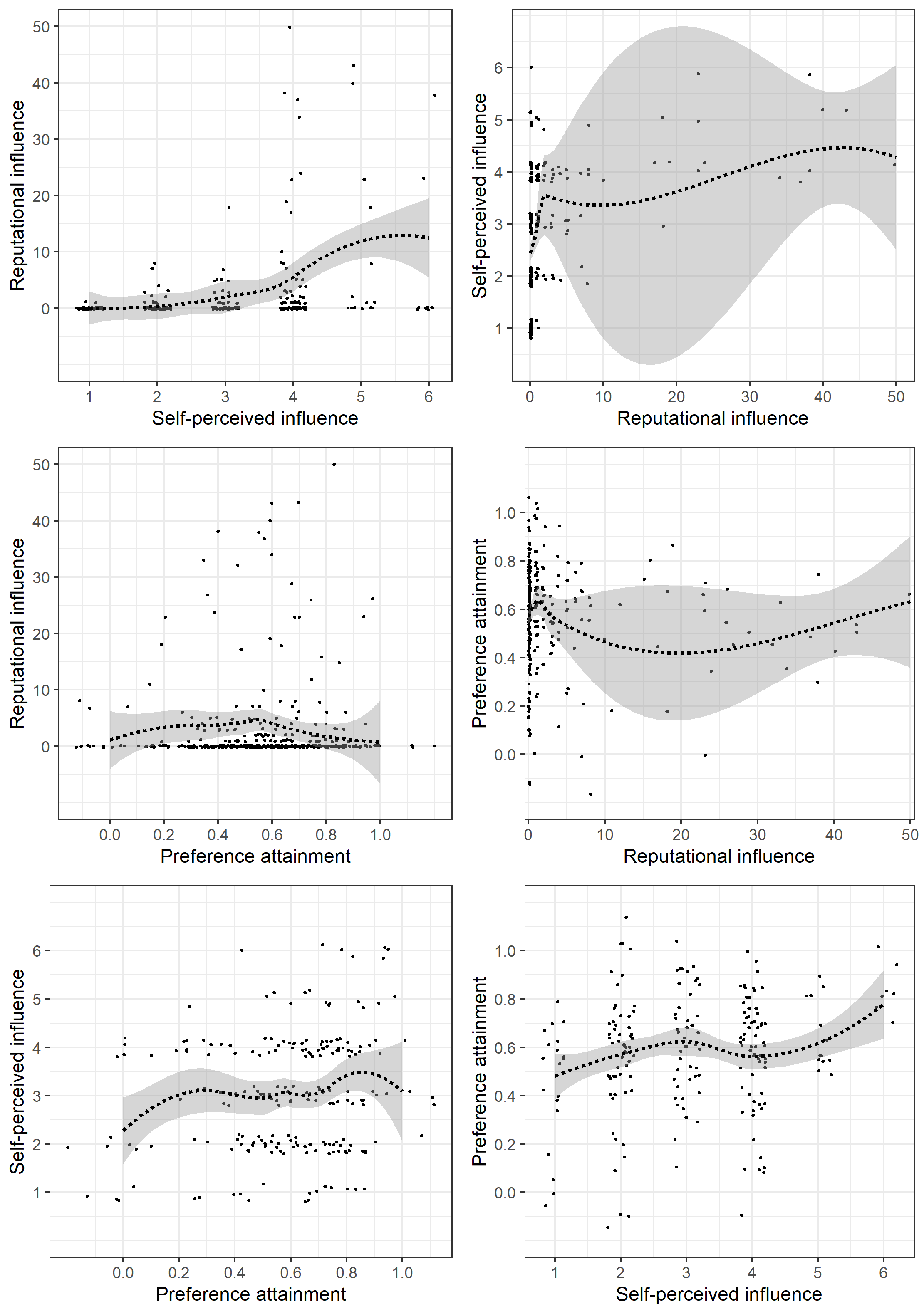}
\caption{Relationship between dependent variables.}
\label{f_influence.measures}
\floatnote{Each row features one of the three combinations of measures. The second column presents the same combination as the first with switched axes. Since the fitted model (Loess regression) attributes all the error to the respective dependent variable, switching the axes results in distinct smoothing lines and confidence intervals. In contrast to the subject-item observations used in the following analyses, in this graph, \textit{preference attainment} is represented on the subject level as the ratio of preferences attained. To make the graphs easier to interpret, minimal noise (0.2 of the resolution of the data in width and height) was added to avoid overlapping observations. Also, the observations for \textit{reputational influence} are  plotted only for values up to 50. This leads to 6 observations (values between 57 and 118) not being featured in the respective graphs. However, the smoothing lines and their confidence intervals are based on the full sample.}
\end{figure}

To begin, we look at the relationship between the dependent variables. The scatterplots in Figure \ref{f_influence.measures} reveal the degree of association between the three variables. The different units of observation (subject level for perceptional measures, subject-item level for \textit{preference attainment}) make it difficult to compare. To account for this, we calculated an additional variable reflecting the ratio of attained preferences, representing \textit{preference attainment} on the subject level. Given the unbalanced nature of the data, the denominator varied according to the number of items the organization indicated their preferences on.

The smoothing lines do hint at some modest level of agreement between \textit{self-perceived} and \textit{reputational influence}. There are short segments with a negative slope. For most observations though, higher values on one measure correspond with higher values on the other. A very different picture emerges from the scatterplots with \textit{preference attainment}. With \textit{reputational influence} there seems to be no common trend at all. In the case of \textit{self-perceived influence}, observations with very high values also tend to exhibit above average values in \textit{preference attainment}. However, the smoothing lines feature major segments of negative slopes or sideward scattering.

\clearpage
\subsection{Regression results}
The descriptive results suggest that the three dependent variables do measure somewhat distinct concepts. This begs the question to what extent this is also reflected in their respective relationships with \textit{political budget}.

Indeed, the regressions in Table \ref{t_reg.compare} reveal differing patterns. In the case of \textit{self-perceived influence}, \texttt{political budget} is a positive and statistically highly significant predictor. A larger \texttt{political budget} is associated with organizations considering themselves more influential. 

The result for the \textit{reputational influence} model are similar. While the top three coefficients in the table refer to the count model, the bottom three feature the results for the zero-model. Both are statistically significant, but the two coefficients for \texttt{political budget} are seemingly contrasting. This is due to the fact that the zero-model predicts zeros, i.e., non-occurrence. Hence, a larger \textit{political budget} corresponds with a decrease in the probability for an organization to be mentioned by none of the survey participants (zero-model). Similarly, the expected number of mentions as an influential actor increases with a larger \textit{political budget} (count-model).

In contrast, the mixed model for the \textit{preference attainment} measure suggests a negative relationship with \textit{political budget}. A larger \textit{political budget} is associated with a lower probability to attain policy preferences. This negative association is statistically different from zero, too, although it does not reach the same level of statistical significance as the estimates for the two perceptional measures of lobbying influence.

Interestingly, organizations representing \textit{business interests} consider themselves less influential and are less likely to attain their preferences. Both estimates are statistically different from zero. Also, the probability to be mentioned as influential by at least one survey participant is lower for such interests. However, this relationship does not reach any common level of statistical significance. 

Although there is substantial overlap, the samples on which these three models build are not identical. For each measure, we included all available observations, which varies between measures. However, the results of all three models were largely the same for a subsample of 182 organizations for which there was data on all three measures (see models 1-3 in Table \ref{app_t_reg.compare.all} in appendix \ref{app_regression}).

As outlined above, the number of items on which the \textit{preference attainment} measure builds varies between participants. Organizations with very few indicated preferences may introduce a potential bias for the \textit{preference attainment} variable. However, the results, again, were robust to the exclusion of 88 subject-item observations from the 24 organization for which there were only 10 observations or fewer available (see model 2 in Table \ref{app_t_reg.compare.pa} in \ref{app_regression}).

Finally, it is important to point out that the effect sizes between the three models are hardly comparable. While the predictor variables are the same, the scale of measurement varies between the three models, rendering a comparison of estimates difficult. Given the different regression models, the same caveat also applies to the question to what extent the specification can explain the variances in the dependent variables. Hence, we also refrain from comparing goodness of fit statistics between the models.

\begin{table}[!h]\centering\footnotesize
\begin{threeparttable}
\renewcommand{\arraystretch}{1.2}
\begin{tabular}{l c c c}
 & \multicolumn{1}{c}{Self-perceived} & \multicolumn{1}{c}{Reputational} & \multicolumn{1}{c}{Preference} \\
 & influence & influence & attainment \\
\hline
Intercept                     & $2.26^{***}$ & $0.17$        & $0.79^{***}$ \\
                              & $(0.15)$     & $(0.15)$      & $(0.17)$     \\
Political budget              & $0.23^{***}$ & $0.45^{***}$  & $-0.05^{**}$ \\
                              & $(0.04)$     & $(0.02)$      & $(0.03)$     \\
Business interest                      & $-0.26^{*}$  & $0.00$        & $-0.23^{**}$ \\
                              & $(0.14)$     & $(0.07)$      & $(0.11)$     \\
Intercept (zero model)        &              & $1.96^{***}$  &              \\
                              &              & $(0.34)$      &              \\
Political budget (zero model) &              & $-0.44^{***}$ &              \\
                              &              & $(0.08)$      &              \\
Business interests (zero model)         &              & $0.50$        &              \\
                              &              & $(0.31)$      &              \\
\hline
R$^2$                         & $0.15$       & $ $            & $ $           \\
Adj. R$^2$                    & $0.14$       & $ $            & $ $           \\
Num. obs.                     & $261$        & $269$         & $4200$       \\
AIC                           & $ $           & $1625.30$     & $5201.85$    \\
Log Likelihood                & $ $           & $-806.65$     & $-2595.92$   \\
BIC                           & $ $           & $ $            & $5233.56$    \\
Num. groups: organization     & $ $           & $ $            & $187$        \\
Num. groups: item             & $ $           & $ $            & $41$         \\
Var: organization (Intercept) & $ $           & $ $            & $0.26$       \\
Var: item (Intercept)         & $ $           & $ $            & $0.61$       \\
\hline
\end{tabular}
\begin{tablenotes}[flushleft]
\scriptsize{\item[\hspace{-5mm}] $^{***}p<0.01$; $^{**}p<0.05$; $^{*}p<0.1$. \item[\hspace{-5mm}] \textit{Note}: Self-perceived influence is estimated using OLS;
                       the model for reputational influence uses maximum likelihood with a poisson regression for the count and a binomial regression with logit link for the zero model;
                       preference attainment is estimated using a mixed model fit by maximum likelihood with random intercepts for organizations and items.}
\end{tablenotes}
\end{threeparttable}
\caption{Regression results - comparison of influence measures.}
\label{t_reg.compare} 
\end{table}

\subsection{Mediation results}

Having outlined the implications the choice of the dependent variable holds, we can now move from our methodological to our final, more substantive contribution. As mentioned above, in most cases, simply holding large financial resources is not going to result in influence over policy automatically. Hence, given the effects of \textit{political budget} on influence identified in the previous analysis, we investigate mechanisms through which such resources may manifest their influence potential. The analysis tests the extent to which 13 different lobbying activities mediate the effect of \textit{political budget} on the influence variables.

\begin{figure}[hbt!]
\includegraphics[width=.93\textwidth]{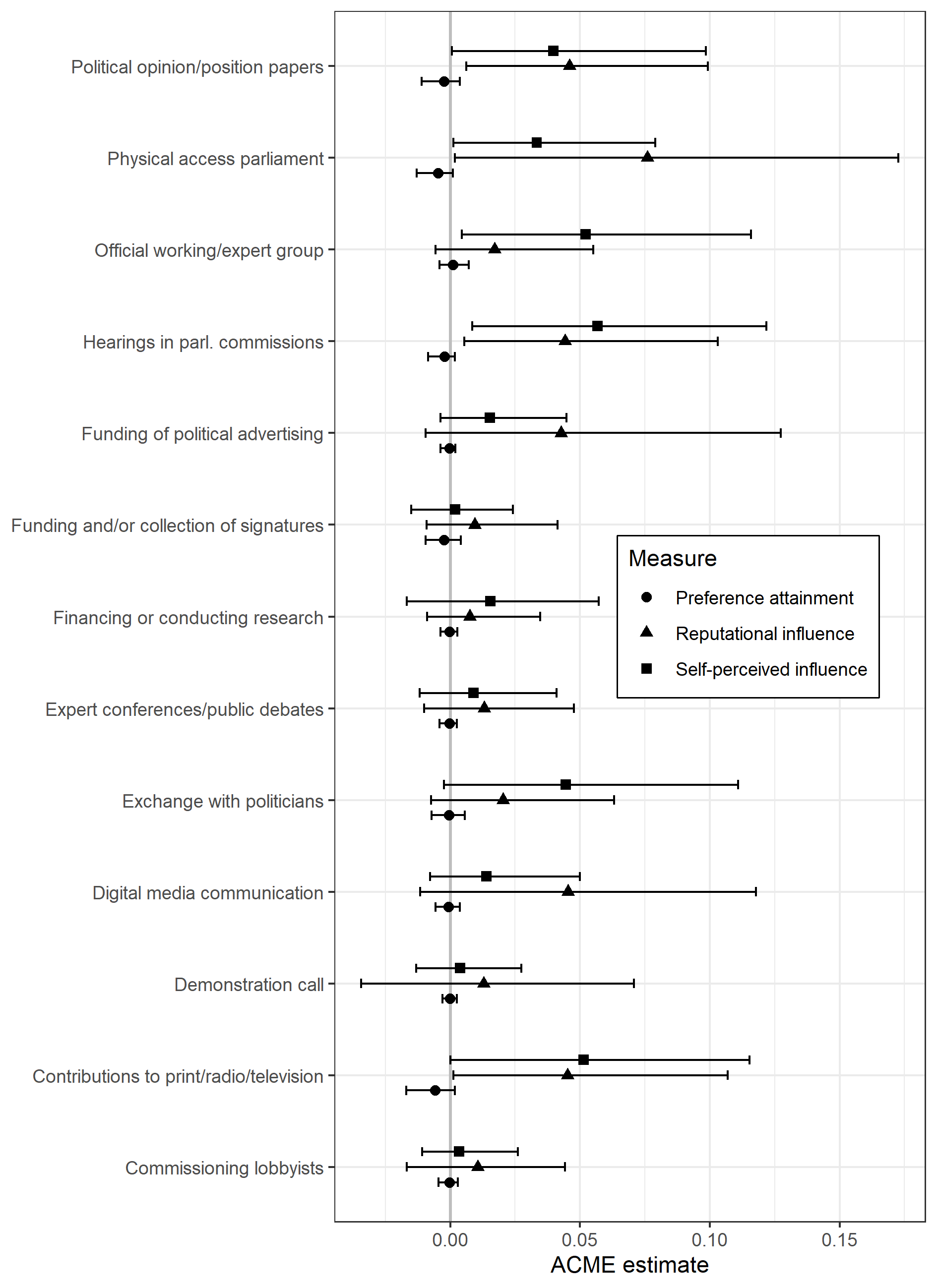}
\caption{Comparison of ACME estimates.}
\label{f_coefplot.bayes}
\floatnote{95\% confidence intervals estimated using quasi-Bayesian Monte Carlo approximation.}
\end{figure}

Figure \ref{f_coefplot.bayes} illustrates the average causal mediation effect (ACME) for each activity. While the figure includes the corresponding confidence intervals, a more comprehensive overview of the mediation results can be found in Table \ref{t_mediation.bayes}\footnote{See Appendix \ref{app_sensitivity} for a discussion of the underlying identification assumptions for the mediation analysis in the context of our analysis.}.

Not all ACMEs reach statistical significance and, more interestingly, there are differences between influence measures. The drafting of political opinion papers, physical access to the Swiss parliament and having participated in parliamentary hearings are statistically significant mediators for both, \textit{self-perceived} and \textit{reputational influence}. Contributions to media outlets mediate the effect of \texttt{political budget} only on \textit{reputational} but not \textit{self-perceived influence}, while the opposite is true in the case of organizations that indicated to have participated in official working groups to draft new legislation. 

Our mediation analysis is not limited to positive effects; i.e., causal mechanisms can relate to negative effects as well. Significant mediation effects in that case would suggest that a larger \textit{political budget} leads organizations to choose activities, which -- perhaps counter-intuitively -- have a negative effect on their lobbying influence. However, the mediation analysis on the \textit{preference attainment} model does not provide any evidence for this to be the case. None of the ACMEs are statistically significantly different from zero.

For both perceptional measures, the results suggest partial mediation as opposed to full mediation for all statistically significant activities. In terms of effect size, relative comparisons between activities seem best suited to clarify the effects, especially given the ordinal nature of the variable used to measure \textit{political budget}. For \textit{reputational influence}, the coefficients for hearings in parliamentary commission, drafting political opinion or position papers and providing contributions to mass media are all comparable in size. Physical access to parliament, however, even after accounting for differences in the ATEs (average treatment effect; aggregated effects of direct and mediated effect), mediates a substantially larger proportion of the effect compared to the other activities. 

For \textit{self-perceived influence}, participation in hearings in parliamentary commission features as the most important mechanism linking \texttt{political budget} with influence, followed by participation in official working or expert groups to draft legislation. Physical access to parliament and drafting political opinion or position papers feature a substantially smaller mediation effect.   


The different scales of measurement prohibit a direct comparison of effect sizes across the different measures of influences. However, some relative inferences can be drawn, nonetheless. While the absolute values are uninformative, the ratio between the coefficients provides further evidence on the importance of physical access to parliament for \textit{reputational influence}. In contrast, for \textit{self-perceived influence}, the relative comparison highlights the relevance of hearings in parliamentary commissions as a mediator.

\begin{landscape}
\begin{table}[!h]\centering
\centering
\resizebox{\linewidth}{!}{
\renewcommand{\arraystretch}{1.2}
\begin{tabular}[t]{>{\raggedright\arraybackslash}p{12em}ccccccccccccccc}
\multicolumn{1}{l}{\em{ }} & \multicolumn{5}{l}{\em{Self-perceived influence}} & \multicolumn{5}{l}{\em{Reputational influence}} & \multicolumn{5}{l}{\em{Preference attainment}} \\
\cmidrule(l{3pt}r{3pt}){2-6} \cmidrule(l{3pt}r{3pt}){7-11} \cmidrule(l{3pt}r{3pt}){12-16}
Mediator & ATE & ADE & ACME & CI 95\% ACME & Mediated & ATE & ADE & ACME & CI 95\% ACME & Mediated & ATE & ADE & ACME & CI 95\% ACME & Mediated\\
\midrule
Exchange with politicians & 0.249 & 0.204 & 0.044 & {}[-0.002, 0.111] & 0.167 & 0.348 & 0.328 & 0.020 & {}[-0.007, 0.063] & 0.053 & -0.009 & -0.009 & -0.001 & {}[-0.007, 0.006] & 0.034\\
Expert conferences/public debates & 0.232 & 0.223 & 0.009 & {}[-0.012, 0.041] & 0.026 & 0.334 & 0.321 & 0.013 & {}[-0.010, 0.048] & 0.034 & -0.009 & -0.009 & 0.000 & {}[-0.004, 0.003] & 0.012\\
Funding of political advertising & 0.211 & 0.195 & 0.015 & {}[-0.004, 0.045] & 0.064 & 0.306 & 0.263 & 0.043 & {}[-0.009, 0.127] & 0.124 & -0.008 & -0.008 & 0.000 & {}[-0.004, 0.002] & 0.015\\
Funding and/or collection of signatures & 0.228 & 0.226 & 0.002 & {}[-0.015, 0.024] & 0.002 & 0.334 & 0.325 & 0.010 & {}[-0.009, 0.041] & 0.020 & -0.008 & -0.005 & -0.002 & {}[-0.009, 0.004] & 0.280\\
Demonstration call & 0.229 & 0.225 & 0.004 & {}[-0.013, 0.027] & 0.010 & 0.333 & 0.320 & 0.013 & {}[-0.034, 0.071] & 0.035 & -0.009 & -0.009 & 0.000 & {}[-0.003, 0.003] & 0.003\\
Commissioning lobbyists & 0.230 & 0.226 & 0.003 & {}[-0.011, 0.026] & 0.006 & 0.334 & 0.323 & 0.011 & {}[-0.017, 0.044] & 0.026 & -0.009 & -0.008 & 0.000 & {}[-0.004, 0.003] & 0.012\\
Physical access parliament & 0.217 & 0.184 & 0.033 & {}[0.001, 0.079] & 0.140 & 0.339 & 0.263 & 0.076 & {}[0.002, 0.173] & 0.216 & -0.008 & -0.003 & -0.005 & {}[-0.013, 0.001] & 0.493\\
Hearings in parl. commissions & 0.208 & 0.151 & 0.057 & {}[0.009, 0.122] & 0.260 & 0.321 & 0.277 & 0.044 & {}[0.005, 0.103] & 0.131 & -0.008 & -0.006 & -0.002 & {}[-0.009, 0.002] & 0.198\\
Official working/expert group & 0.226 & 0.174 & 0.052 & {}[0.005, 0.116] & 0.218 & 0.334 & 0.317 & 0.017 & {}[-0.006, 0.055] & 0.045 & -0.009 & -0.010 & 0.001 & {}[-0.004, 0.007] & -0.053\\
Political opinion/position papers & 0.238 & 0.199 & 0.040 & {}[0.001, 0.098] & 0.150 & 0.329 & 0.283 & 0.046 & {}[0.006, 0.099] & 0.135 & -0.010 & -0.007 & -0.002 & {}[-0.011, 0.004] & 0.194\\
Financing or conducting research & 0.229 & 0.214 & 0.015 & {}[-0.017, 0.057] & 0.058 & 0.331 & 0.324 & 0.008 & {}[-0.009, 0.035] & 0.016 & -0.009 & -0.009 & 0.000 & {}[-0.004, 0.003] & 0.005\\
Contributions to print/radio/television & 0.233 & 0.182 & 0.051 & {}[-0.000, 0.115] & 0.208 & 0.334 & 0.289 & 0.045 & {}[0.001, 0.107] & 0.128 & -0.010 & -0.004 & -0.006 & {}[-0.017, 0.002] & 0.535\\
Digital media communication & 0.229 & 0.215 & 0.014 & {}[-0.008, 0.050] & 0.050 & 0.349 & 0.303 & 0.045 & {}[-0.011, 0.118] & 0.124 & -0.009 & -0.008 & -0.001 & {}[-0.006, 0.004] & 0.033\\
\bottomrule
\multicolumn{16}{l}{\rule{0pt}{1em}\textit{Note: } 95\% confidence intervals estimated using quasi-Bayesian Monte Carlo approximation.}\\
\end{tabular}}
\caption{Mediation analysis estimates.}
\label{t_mediation.bayes} 
\end{table}
\end{landscape}

\subsection{Inside vs. outside lobbying}

A closer look at the mediation results reveals an interesting pattern. In the case of \textit{self-perceived influence}, it is mainly inside lobbying activities that appear to translate the effect of \textit{political budget}. Physical access to parliament, participation in official working or expert groups and hearings in parliamentary commissions all aim to influence decision-makers directly. Perhaps, actors may exploit political position and opinion papers in the context of an outside lobbying strategy too. Typically, however, they serve to inform and persuade insiders of the policymaking process.

For \textit{reputational influence}, the classification is less pronounced. Contributing to print, radio, or television, one of the significant mediating activities, represents an outside lobbying strategy. Nonetheless, in addition to the unclassifiable activity of drafting political position or opinion papers, both other significant mediators represent typical elements of an inside lobbying strategy.

\begin{figure}[hbt!]
\includegraphics[width=.85\textwidth]{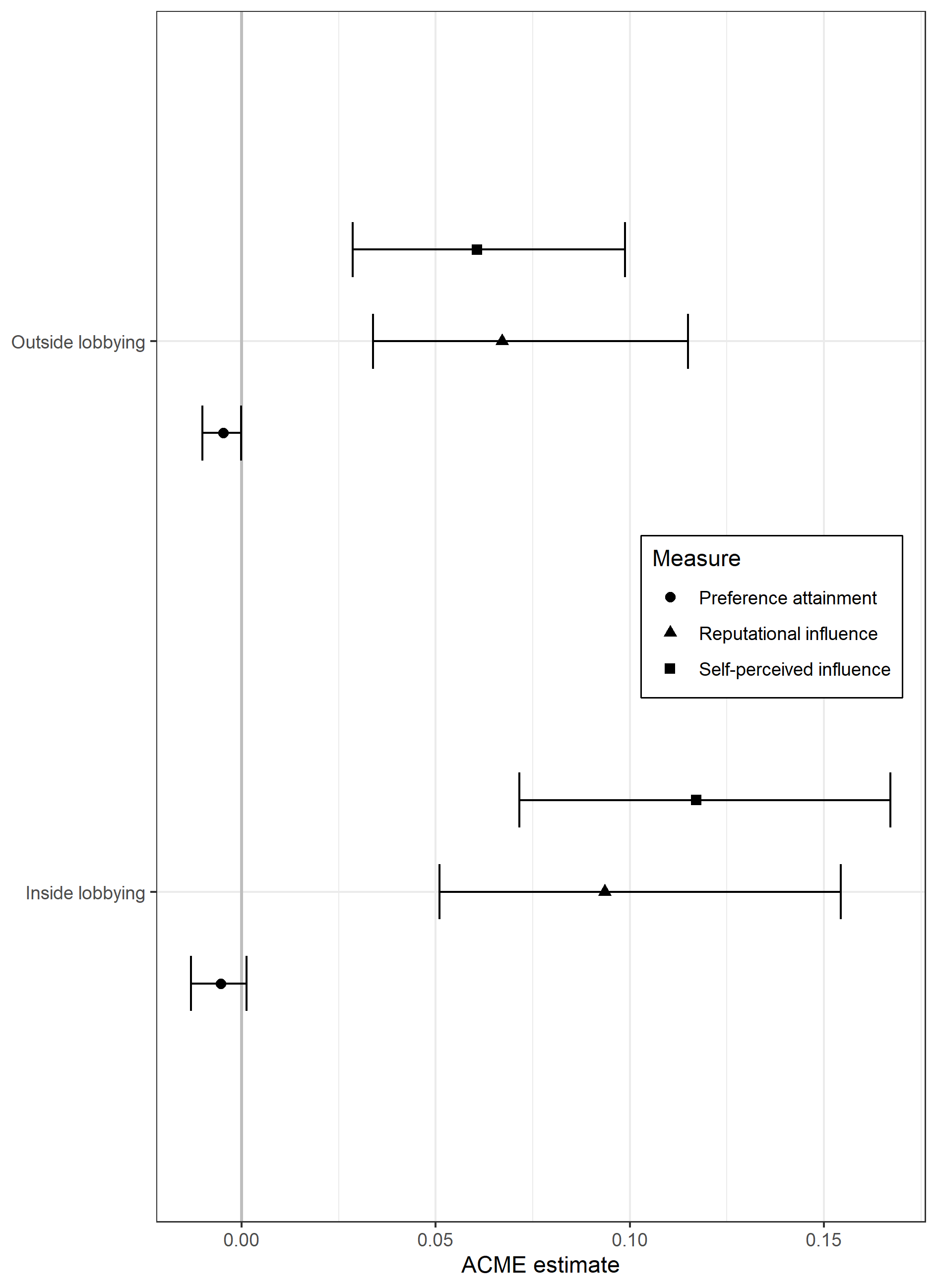}
\caption{Comparison of ACME estimates - inside vs. outside lobbying.}
\label{f_coefplot.insideoutside}
\floatnote{95\% confidence intervals estimated using quasi-Bayesian Monte Carlo approximation.}
\end{figure}

Hence, there is reason to suspect a fundamental difference between inside and outside strategies in terms of their ability to translate \textit{political budget} into influence. In the following, final analysis, we put this alleged dichotomy to test (see Table \ref{t_mediation.insideoutside} and Figure  \ref{f_coefplot.insideoutside}). Aggregating the individual, binary activity variables into two overall inside and outside lobbying variables, we rerun the analysis using the two new variables as mediators \footnote{Our classification of inside and outside lobbying activities builds on \cite{Binderkrantz2005a}. We assigned "Exchange with politicians", "Commissioning lobbyists", "Physical access to parliament", "Hearings in parl. commissions", "Official working/expert group", "Political opinion/position papers" to the inside lobbying category and "Financing or conducting research", "Contributions to print/radio/television", "Digital media communication", "Expert conferences/public debates", "Funding of political advertising", "Funding and/or collection of signatures",  "Demonstration call" to the outside lobbying category.}.

\begin{table}[!h]\centering\footnotesize
\renewcommand{\arraystretch}{1.2}
\begin{tabular}[t]{p{.1\linewidth}lcccccc}
 & Mediator & ATE & ADE & ACME & CI 95\% ACME & Mediated\\
\midrule
\multicolumn{7}{l}{\textit{Self-perceived influence}}\\
& Inside lobbying & 0.230 & 0.113 & 0.117 & {}[0.072, 0.167] & 0.503\\
& Outside lobbying & 0.232 & 0.171 & 0.061 & {}[0.029, 0.099] & 0.260\\
\multicolumn{7}{l}{\textit{Reputational influence}}\\
& Inside lobbying & 0.328 & 0.235 & 0.093 & {}[0.051, 0.154] & 0.282\\
& Outside lobbying & 0.328 & 0.261 & 0.067 & {}[0.034, 0.115] & 0.200\\
\multicolumn{7}{l}{\textit{Preference attainment}}\\
& Inside lobbying & -0.009 & -0.004 & -0.005 & {}[-0.013, 0.001] & 0.556\\
& Outside lobbying & -0.009 & -0.004 & -0.005 & {}[-0.010, -0.000] & 0.483\\
\bottomrule
\multicolumn{7}{l}{\rule{0pt}{1em}\textit{Note: } 95\% confidence intervals estimated using quasi-Bayesian Monte Carlo approximation.}\\
\end{tabular}

\caption{Mediation analysis estimates - inside vs. outside lobbying}
\label{t_mediation.insideoutside} 
\end{table}

In contrast to the findings from the individual activities, the difference between inside and outside lobbying strategies for the aggregated variables is less clear. For both, \textit{self-perceived} and \textit{reputational influence}, the effect size of the aggregated inside lobbying variable is substantially larger than for outside lobbying. However, not only the inside but also the outside mediator variable reached statistical significance for both perceptional measures.

Interestingly, the analysis of the aggregated mediator variables revealed another interesting finding with regard to the \textit{preference attainment} measure. As shown above, we found that none of the individual lobbying activities mediate the negative effect of \textit{political budget} on \textit{preference attainment} in a statistically significant manner. In contrast, the aggregated outside lobbying variable does. Moreover, in proportion to the overall effect of \textit{political budget} on \textit{preference attainment}, the proportion mediated is substantial.

\subsection{Robustness checks}
In the appendix we present the results for several robustness checks. The three models in Table \ref{app_t_reg.compare.all} use the same specifications as the main regressions. However, the sample is limited to the 182 observations for which there are values for all three measures. The results as well as the statistical significance remain largely unchanged. 

Given the numerous contributions in the literature that employ alternative measures for financial resources, in Table \ref{app_t_reg.compare.polstaff} we present the results for the three dependent variables using the number of full time equivalent employees that are commissioned to follow political events rather than \textit{political budget}. The sample sizes are slightly larger. With respect to the perceptional measures of lobbying influence, the results for political staff are similar to those obtained when using \texttt{political budget} in the main analysis, but there is no statistically significant association with \textit{preference attainment}. However, the explanatory power of \texttt{business interests} increases with the use of \texttt{political staff} for the \textit{preference attainment} model. The same is true for \textit{self-perceived influence}.

Table \ref{app_t_reg.compare.budget} features the results for \texttt{overall budget} instead of \texttt{political budget}. The relationship between budget and influence is insignificant for the \textit{self-perceived influence} measure and \textit{preference attainment}. At the same time, the estimates in the \textit{reputational} model increase substantially.

Model 1 of Table \ref{app_t_reg.compare.pa}, features the estimates for the \textit{preference attainment} regression that uses only one random intercept for the grouping variable organization. We find that the estimate for \texttt{political budget} remains statistically significant although at a lower level. In Model 2 we return to the normal specification but only include organizations for which there are more than ten subject-item observations. Results are very similar to the ones in the main regression, including the statistical significance.

In Table \ref{app_t_reg.compare.pa_indiv} we present the results for the subsamples from the two consultations. Not only are the samples smaller than in the main analysis, but the sample size also varies substantially between the two models. While the estimates for \texttt{political budget}  change only little, statistical significance decreases in both cases. Interestingly, in the case of the 'Energiestrategie 2050', the coefficient for Business interests doubles in size compared to the main regression. It also gains statistical significance, while the opposite is true for the consultation on 'Klimapolitik nach 2020'.

Finally, we also conducted the same mediation analysis using bootstrapped rather than Bayes confidence intervals in Figure \ref{app_f_coefplot.boot} and Table \ref{app_t_mediation.boot}. For \textit{reputational influence}, the same activities that were identified to mediate in the main analysis were also found to have statistically significant indirect effect with this alternative confidence interval estimation procedure. In the case of \textit{self-perceived influence}, in addition to the activities from the main analysis, exchange with politicians and media contributions also featured estimates that are statistically significantly different from zero.

\section{Discussion}
We began our empirical analysis by asking whether the three measures of political influence behave in similar manners. Note that conceptually, preference attainment places greater emphasis on observed policy outputs (hence it allows us to construct an objective score), whereas self-perceived influence and reputational influence place more emphasis on actors’ agency in shaping policies, which is an unobserved dynamic stemming from the actors’ intention to shape policies (hence these approaches force us to rely on their stated perceptions). 

Comparing our three measures based on a nested set of actors from a uniform political context, three noteworthy patterns emerge. First, these different conceptual emphases behind preference attainment and the two perception-based measures indeed lead to distinct scoring. The finding indicates that preference attainment measures a different construct from the other two. Second, studies using preference attainment as a measure of political influence would lead to the conclusion that stakeholders with greater financial resources are less likely to achieve their policy goals. In contrast to this finding, our analysis confirms the popular expectation that a larger political budget is indeed more likely to lead to greater political influence when the influence is operationalized by the other two measures, self-perceived and reputational influence. Finally, our results offer empirical evidence for the similarity between two perception-based measures. Self-perceived influence and reputational influence do not only correlate positively with each other, but also correlate positively with political budget. 

In the first place, we wanted to compare the three measures because prior literature has often used them interchangeably as proxies for political influence. Our findings suggest, however, that it is important for researchers to acknowledge the difference both theoretically and empirically. In our view, preference attainment is a measure of policy goal achievement, which is a convolution of \emph{all} the actors’ policy preferences and political influence. The measure cannot indicate a single actor’s influence. The similarity between self-perceived and reputational measures implies that an organization’s self-assessment of its political influence is not detached from how the other actors assess its influence. This is an assuring result for researchers using either of the measures. Here one might eventually consider a practical design element. The self-perceived measure would provide complete information about political influence  for the entire sample of actors, whereas using the reputational measure, the data tend to include many more actors that are never mentioned by the other actors as influential. We cannot capture the potential variation among actors that are not highly prominent. 

In the multivariate analysis, we also considered the role of \emph{business interests} in predicting political influence. This variable measuring whether an organization represents business interests has frequently appeared in prior studies \citep[e.g.,][]{Yackee2006ABureaucracy,Dur2019TheUnion}. While it is often associated with a larger \textit{political budget} \citep[e.g.,][]{McKay2012,Varone2020}, in some cases it has even been used as a proxy for financial resources \citep[e.g.,][]{Binderkrantz2014AConsultations}. Interestingly, in our analyses \emph{business interests} were associated with less influence for \textit{self-perceived influence} as well as \textit{preference attainment}. For the zero model of \textit{preference attainment}, there is a similar trend that does not reach statistical significance, however. 
Hence, against common belief and numerous prior contributions on the topic, our analysis hints at a reduced responsiveness of policymakers for \emph{business interests} \citep[e.g.,][]{Gilens2014TestingCitizens,Balles2020SpecialAttention,Varone2020}. 

Finally, we turn to the mechanism through which \emph{political budget} might affect lobbying influence. Prior literature has theorized that financial resources allow stakeholders to engage in various lobbying activities, and these activities in turn increase the chance that the actors influence policies. By explicitly modeling the link between \emph{political budget} and \emph{influence} that is mediated by various lobbying strategies, we found several important patterns that help to solve the puzzles posited by existing studies. First, both inside and outside lobbying strategies matter. More precisely, financial resources that are materialized through either inside lobbying strategies (activities that are directly targeted at policymakers) or outside lobbying strategies (activities that create similar electoral pressure indirectly by influencing the general public) would increase the chance for the actor to influence policies. The pattern is consistent between the self-perceived and reputational influence measures. Some prior work has also noted that inside lobbying activities, i.e., direct access to politics or politicians, is a privilege that some, but not all,  actors enjoy. Behind this argument is a theoretical conjecture that inside lobbying is more effective in influencing policy. Our findings hint at this pattern. For the self-perceived and reputational influence measure respectively, the effect of financial resources that is transmitted through inside lobbying practices exhibits a larger effect than that through outside lobbying practices. 

Before concluding this paper, we would like to come back to the distinct behavior of the preference attainment measures. As can be seen in Figure~\ref{f_coefplot.insideoutside}, the mediation analysis again shows that \emph{political budget} is negatively associated with \emph{political influence}, and the pattern holds equally through inside and outside lobbying strategies. 
There are very different explanations for the negative effect of \emph{political budget} on lobbying influence. Taken at its face value, the results suggest that actors with a larger \textit{political budget} attain fewer preferences. Moreover, the mediation analysis suggests that these actors engage in a high number of outside activities and that such a lobbying strategy leads to fewer preferences attained. Consequently, it is possible that primarily the use of an outside lobbying strategy leads to lower preference attainment. Thus, a lager \textit{political budget} would not necessarily be associated with fewer  preferences attained. Rather, a larger \textit{political budget} could lead to the use of more outside lobbying strategies, which would then have a negative effect on preference attainment. The fact that a comparatively high share of the effect of \textit{political budget} on preference attainment appears to run through the outside count variable lends support to this hypothesis. However, it should be noted that in the literature, the use of an outside lobbying strategy is typically associated with low-budget campaigns. Wealthy actors supposedly use their resources for preferential access to decision makers and thus are more likely to pursue an inside lobbying strategy.

Alternatively, the results may stem from the fact that the chosen application of preference attainment defines the window of influence taking too narrowly. It is conceivable, for example, that the first draft of the bills already takes into account the preferences of financially powerful actors to the maximum extent. Such regulatory capture of the administrative authorities has already been described in detail in the literature. Changes made in the subsequent policy process in parliament can then only negatively influence the preference attainment score of such actors. As a consequence, the presented analysis would truthfully identify a negative association between \textit{political budget} and attained preferences between the first bill and the final parliamentary bill. However, using it as an influence measure would misrepresent the influence of \textit{political budget} on energy policy overall. This is in contrast to the perceptional measures, which can also take into account possible influence prior to the first bill.

\newpage
\section{Conclusion}
Understanding how interest groups shape policy has long been of interest to many political scientists and economists. However, various empirical findings have not been in agreement as to even the most (seemingly) straightforward question concerning lobbying influence: whether better-endowed stakeholders influence policies more. 
Our paper revisited this relationship between financial resources and lobbying influence, in such a way that would help us understand potential sources of discrepancies. 

We approached this problem from three angles. First, carefully studying varying constructs of lobbying influence employed in the literature, we built our analysis on a comparison of three types of influence measures that appear frequently in the existing studies—the \textit{self-perceived}, \textit{reputation-based}, and \textit{preference attainment} measures. What is important here is that we constructed the three measures based on the same (or nested) set of actors by combining data from an original survey with 312 Swiss energy policy stakeholders with document data from policy consultation processes. Existing empirical studies employ different measures of influence in different policy contexts, but few scholars seem to be concerned that the qualitatively different conclusions might simply be artifacts of distinct influence measures (and the selection of other covariates). 

Second, as our survey directly probed the size of budget allocated specifically to political purposes, it allowed us to overcome common challenges associated with the operationalization of relevant financial resources. Finally, we addressed the issue of potential confounding between actors’ \textit{political budget} and various lobbying activities. In other words, we tested whether \textit{political budget} can be at work by enabling various lobbying activities, instead of treating their effects independently as the literature has suggested. 

Public policymaking provides ample opportunity for stakeholders to exert influence on legislation and this may well be intended. Transparency on who successfully influences policy as well as common knowledge of the effectiveness of resources or strategies may represent important complements, however.
Our study reveals that the identification of influential actors strongly depends on the influence measure applied. Similarly, conclusions on the effect of \textit{political budget} on influence may be comparable among perceptional measures of influence but contrast with the findings from the \emph{preference attainment} measure. Finally, there is much to suggest that the lobbying activities that an actor engages in, represent a crucial factor in translating the potential of \textit{political budget} into actual influence.

Regarding future research, observing the relationships from this analysis over time represents the logical next step. As in most studies of influence, our empirical strategy cannot rule out the possibility of reverse causality completely. It is possible that the influence of an organization, or the perception thereof, leads to resources being directed towards such actors. Hence, a longitudinal research design may be able to address such concerns. However, given the difficulty of obtaining suitable data even for one point in time only, such an endeavor would require substantial resources.

Nonetheless, such an approach may also be able to further corroborate some of the interpretations deduced from the mediation analyses. Indeed, introducing within-actor variance in terms of lobbying activities as well as \textit{political budget} may further substantiate the plausibility of the underlying, untestable assumptions.

Finally, it would be interesting to learn to what extent our findings also extend to other institutional environments. For example, in addition to the lobbying taking place in parliament, the \textit{preference attainment} scores may be susceptible to the exertion of influence before the authorities submit the first draft to the official consultation process. A comparison between or within countries, taking into account differences in the policy process, may serve to highlight the robustness, or lack thereof, of our findings on the role of \textit{political budget} and lobbying activities regarding lobbying influence.

\newpage
\bibliography{ref_directMendeley.bib}

\clearpage
\appendix
\renewcommand\thetable{\thesection.\arabic{table}}
\renewcommand\thefigure{\thesection.\arabic{figure}}
\setcounter{table}{0}
\setcounter{figure}{0}
\section{Appendix A}\label{app_consultation}
\renewcommand{\arraystretch}{.8}
\footnotesize
\begin{tabularx}{.96\textwidth}{ c X }
\caption{Items asked in consultation for 'Energiestrategie 2050'.}
\label{app_t_consultation.item.es2050}\\
\toprule\endfirsthead
\toprule\endhead
\bottomrule\endlastfoot
No.     & Description \\
\midrule
1 & Do you agree overall with the consultation draft on the Energy Strategy 2050?\\
2 & Do you agree with the staged approach of the Energy Strategy 2050 (second stage according to section 1.4 in the explanatory report)?\\
3 & Do you agree with linking the gradual phase-out of nuclear energy to the present package of measures?\\
4 & Do you agree that general licenses for the construction of new nuclear power plants should no longer be issued?\\
5 & Do you agree that expansion targets for the production of electricity from renewable energies as well as consumption targets should be set by law?\\
6 & Do you agree with the envisaged increase in total federal and cantonal funding to strengthen the building program from 2015 to a maximum of 600 million Swiss francs per year?\\
7 & Which variant do you prefer when changing the legal basis for the use of the proceeds from the CO2 levy for the building sector? \\
8 & Do you agree that costs for building investments that serve to save energy and protect the environment can be tax-deductible spread over three years and that, from 2025, investments (cf. explanatory report: 2.2.3) that serve to save energy and protect the environment will only be tax-deductible if the building concerned meets a certain minimum energy standard?\\
9 & Do you agree with the tightening of the CO2 emissions target for passenger cars first placed on the market to an average of 95 g CO2/km by the end of 2020?\\
10 & Do you agree with the introduction of a CO2 emissions target for vans and light semitrailers first placed on the market and setting it at an average of 175 g CO2/km by the end of 2017 and at an average of 147 g CO2/km by the end of 2020?\\
11 & Do you agree that electricity suppliers have to meet targets to continuously increase the efficiency of electricity consumption (by introducing so-called white certificates)?\\
12 & Do you agree that the federal government can oblige companies in the energy industry to publish data for reasons of transparency and information (in particular regarding electricity and heat consumption by customer groups and regarding offers and measures to promote domestic and renewable energies and energy efficiency)?\\
13 & Do you agree with the extension of competitive bidding to electricity production and distribution?\\
14 & Do you agree that end consumers with an electricity consumption of more than 0.5 GWh per year can commit to the federal government to increase electricity efficiency and reduce CO2 emissions and thus be reimbursed for the grid surcharge?\\
15 & Do you agree with the introduction of joint federal and cantonal planning and a nationwide expansion potential plan for the expansion of renewable energies?\\
16 & Do you agree that the cantons should be obliged to define suitable areas and stretches of water in the structure plan, in particular for hydropower and wind power, and to submit a utilization plan for this purpose?\\
17 & Do you agree that for new renewable energy installations above a certain size and importance, a national interest should be established?\\
18 & Do you agree with the introduction of a self-consumption regulation, i.e. the creation of the legal possibility for plant operators to consume self-produced energy themselves?\\
19 & Do you agree with the exclusion of waste-to-energy plants, sewage treatment plants, and plants that partially use fossil fuels from the group of eligible plants?\\
20 & Do you agree with the limitation of the annual financial resources available for the promotion of photovoltaic systems?\\
21 & Do you agree that a separate body in the form of a subsidiary at the national grid company should be created for the implementation of the feed-in tariff system and the new tasks (one-off tariff for small photovoltaic systems, CHP tariff system)?\\
22 & Do you agree that photovoltaic systems with an output of less than 10 kW are subsidized outside the feed-in tariff model?\\
23 & Do you agree that photovoltaic systems with an output of less than 10 kW should be subsidized with a one-off contribution (one-off payment) instead of the feed-in tariff?~Or do you prefer - as an alternative to the one-off payment - net metering for the future promotion of small photovoltaic systems with an output of less than 10 kW? EnG, Art. 28-30, draft of September 28, 2012 \\
24 & Do you agree that the small photovoltaic plants below 10 kW on the waiting list (without a positive decision) should be excluded from the feed-in tariff system and subsidized by means of a one-off payment?\\
25 & Do you agree with the removal of the overall cap as well as the partial caps on compensation funding?\\
26 & Do you agree with the introduction of a WKK remuneration system?\\
27 & Do you agree with the support range of the remuneration system for CHP (plants with a firing thermal capacity of 0.35 MW up to and including 20 MW)?\\
28 & Do you agree with the introduction of an obligation to offset all emissions caused, with simultaneous exemption from the CO2 tax, for plants participating in the CHP remuneration system?\\
29 & Do you agree with the proposed regulations to speed up proceedings in the field of electricity law? In particular, this includes restricting access to the Federal Supreme Court to legal questions of fundamental importance.\\
30 & Do you agree with the proposed regulations on the introduction and cost-bearing of smart metering systems?\\
\end{tabularx}

\clearpage
\renewcommand{\arraystretch}{.8}
\footnotesize
\begin{tabularx}{.96\textwidth}{ c X }
\caption{Items asked in consultation for 'Klimapolitik nach 2020'.}\\
\label{app_t_consultation.item.klima}\\
\toprule\endfirsthead
\toprule\endhead
\bottomrule\endlastfoot
No.     & Description \\
\midrule
1   & Do you agree in principle with the consultation draft on climate policy after 2020 (Paris Agreement, agreement with the EU on linking the two emissions trading systems, total revision of the CO2 Act)?\\
2   & Should Switzerland ratify the Paris Agreement?\\
3   & Do you agree with the overall target and with the average target for Switzerland?\\
4   & Do you agree with the proposed domestic targets (-30\% by 2030 compared to 1990 and -25\% on average for 2021-2030 compared to 1990)?\\
5   & Do you agree with the linking of the Swiss and EU emissions trading systems?\\
6   & Do you agree with the continuation of the CO2 tax on fuels according to the proven mechanism for increasing the tax depending on the development of emissions and up to the proposed maximum rate of 240 francs per ton of CO2?\\
7   & Do you agree with the continuation of the rule for exemption from the levy for emissions-intensive companies that do not participate in the emissions trading system?\\
8   & Do you agree that the exemption entitlement should be derived from the ratio of the CO2 emission burden of the company to the relevant salary of the employees and that it should be at least 1 percent?\\
9   & Which of the two proposed variants for the design of the levy exemption do you prefer in principle?\\
10  & If you do not fully agree with either of the two proposed options, how do you think the tax exemption mechanism should be designed?\\
11  & Do you agree that the partial earmark for the Buildings Program should be limited until 2025, detached from the KELS bill?\\
12  & Do you agree that a subsidiary ban on the replacement of existing and the installation of new fossil heating systems can be activated in the event that CO2 emissions from buildings are not reduced sufficiently?\\
13  & Do you agree with the exemptions foreseen by the law in case the ban on fossil heating is activated?\\
14  & Do you agree with the continuation of the compensation obligation for importers of fossil fuels, including the proposed division between domestic and foreign compensation?\\
15  & Do you agree with a continuation of the CO2 emission regulations for vehicles (for passenger cars as well as for vans and light semitrailers) in line with the EU?\\
16  & Do you agree with the abolition of annual contributions to the technology fund from 2025 (abolition of partial earmarking of the CO2 levy on fuels) separately from the KELS bill?\\
17  & Do you agree with the continuation of the activities for education and training as well as for information and consultation of the public and the professionals concerned?\\
19  & In your view, are there further reduction measures that the Federal Council should submit to Parliament? If so, which ones?\\
20  & Do you have any other comments on the bill?   
\end{tabularx}

\clearpage
\section{Appendix B}\label{app_regression}

\begin{table}[!h]\centering\footnotesize
\begin{threeparttable}
\renewcommand{\arraystretch}{1.2}
\begin{tabular}{l c c c}
 & \multicolumn{1}{c}{Self-perceived} & \multicolumn{1}{c}{Reputational} & \multicolumn{1}{c}{Preference} \\
 & influence & influence & attainment\\
\hline
\noalign{\vskip 0.1cm}
Intercept                     & $2.48^{***}$ & $0.25^{*}$    & $0.80^{***}$ \\
                              & $(0.16)$     & $(0.15)$      & $(0.17)$     \\
Political budget              & $0.22^{***}$ & $0.45^{***}$  & $-0.05^{**}$ \\
                              & $(0.04)$     & $(0.03)$      & $(0.03)$     \\
Business interests                     & $-0.30^{*}$  & $-0.04$       & $-0.23^{**}$ \\
                              & $(0.16)$     & $(0.07)$      & $(0.11)$     \\
Intercept (zero model)        &              & $1.44^{***}$  &              \\
                              &              & $(0.36)$      &              \\
Political budget (zero model) &              & $-0.38^{***}$ &              \\
                              &              & $(0.09)$      &              \\
Business interests (zero model)         &              & $0.45$        &              \\
                              &              & $(0.34)$      &              \\
\hline
R$^2$                         & $0.16$       & $ $            & $ $           \\
Adj. R$^2$                    & $0.15$       & $ $            & $ $           \\
Num. obs.                     & $182$        & $182$         & $4084$       \\
AIC                           & $ $           & $1483.33$     & $5040.05$    \\
Log Likelihood                & $ $           & $-735.67$     & $-2515.03$   \\
BIC                           & $ $           & $ $            & $5071.63$    \\
Num. groups: organization     & $ $           & $ $            & $182$        \\
Num. groups: item             & $ $           & $ $            & $41$         \\
Var: organization (Intercept) & $ $           & $ $            & $0.27$       \\
Var: item (Intercept)         & $ $           & $ $            & $0.63$       \\
\hline
\end{tabular}
\begin{tablenotes}[flushleft]
\scriptsize{\item[\hspace{-5mm}] $^{***}p<0.01$; $^{**}p<0.05$; $^{*}p<0.1$.
\item[\hspace{-5mm}] \textit{Note}: All regressions use the same observations; i.e., only organizations for which there is a \textit{preference attainment} measure as well as data on self-perceived and \textit{reputational influence}. \textit{Self-perceived influence} is estimated using OLS; the model for \textit{reputational influence} uses maximum likelihood with a poisson regression for the count and a binomial regression with logit link for the zero model; \textit{preference attainment} is estimated using a mixed model fit by maximum likelihood with random intercepts for organizations and items.}
\end{tablenotes}
\end{threeparttable}
\caption{Regression results - comparison of influence measures with equal sample size.}
\label{app_t_reg.compare.all}
\end{table}

\begin{table}[!h]\centering\footnotesize
\begin{threeparttable}
\renewcommand{\arraystretch}{1.2}
\begin{tabular}{l c c c}
 & \multicolumn{1}{c}{Self-perceived} & \multicolumn{1}{c}{Reputational} & \multicolumn{1}{c}{Preference} \\
 & influence & influence & attainment\\
\hline
\noalign{\vskip 0.1cm}
Intercept                     & $2.95^{***}$  & $2.20^{***}$ & $0.69^{***}$  \\
                              & $(0.12)$      & $(0.05)$     & $(0.15)$      \\
Political staff               & $0.03^{**}$   & $0.04^{***}$ & $-0.01$       \\
                              & $(0.01)$      & $(0.00)$     & $(0.01)$      \\
Business interests                     & $-0.43^{***}$ & $0.10$       & $-0.32^{***}$ \\
                              & $(0.15)$      & $(0.07)$     & $(0.11)$      \\
Intercept (zero model)        &               & $0.75^{***}$ &               \\
                              &               & $(0.21)$     &               \\
Political staff (zero model)  &               & $-0.06$      &               \\
                              &               & $(0.04)$     &               \\
Business interests (zero model)         &               & $0.49^{*}$   &               \\
                              &               & $(0.27)$     &               \\
\hline
R$^2$                         & $0.05$        & $ $           & $ $            \\
Adj. R$^2$                    & $0.04$        & $ $           & $ $            \\
Num. obs.                     & $277$         & $286$        & $4555$        \\
AIC                           & $ $            & $2254.01$    & $5565.78$     \\
Log Likelihood                & $ $            & $-1121.01$   & $-2777.89$    \\
BIC                           & $ $            & $ $           & $5597.90$     \\
Num. groups: organisation     & $ $            & $ $           & $198$         \\
Num. groups: item             & $ $            & $ $           & $41$          \\
Var: organisation (Intercept) & $ $            & $ $           & $0.34$        \\
Var: item (Intercept)         & $ $            & $ $           & $0.67$        \\
\hline
\end{tabular}
\begin{tablenotes}[flushleft]
\scriptsize{\item[\hspace{-5mm}] $^{***}p<0.01$; $^{**}p<0.05$; $^{*}p<0.1$.
\item[\hspace{-5mm}] \textit{Note}: Instead of \texttt{Political budget} as the main explanatory variable, the table features the results for the number of FTE commissioned to follow political events. \textit{Self-perceived influence} is estimated using OLS; the model for \textit{reputational influence} uses maximum likelihood with a poisson regression for the count and a binomial regression with logit link for the zero model; \textit{preference attainment} is estimated using a mixed model fit by maximum likelihood with random intercepts for organizations and items.}
\end{tablenotes}
\end{threeparttable}
\caption{Regression results - comparison of influence measures using the size of the political staff to measure financial resources.}
\label{app_t_reg.compare.polstaff} 
\end{table}

\begin{table}[!h]\centering\footnotesize
\begin{threeparttable}
\renewcommand{\arraystretch}{1.2}
\begin{tabular}{l c c c}
 & \multicolumn{1}{c}{Self-perceived} & \multicolumn{1}{c}{Reputational} & \multicolumn{1}{c}{Preference} \\
 & influence & influence & attainment\\
\hline
\noalign{\vskip 0.1cm}
Intercept                     & $2.71^{***}$  & $1.30^{***}$  & $0.59^{***}$ \\
                              & $(0.22)$      & $(0.14)$      & $(0.20)$     \\
Political budget                        & $0.08$        & $0.26^{***}$  & $0.02$       \\
                              & $(0.06)$      & $(0.03)$      & $(0.04)$     \\
Business interests                      & $-0.43^{***}$ & $-0.00$       & $-0.26^{**}$ \\
                              & $(0.16)$      & $(0.07)$      & $(0.11)$     \\
Intercept (zero model)        &               & $1.64^{***}$  &              \\
                              &               & $(0.41)$      &              \\
Political budget (zero model)           &               & $-0.28^{***}$ &              \\
                              &               & $(0.10)$      &              \\
Business interests (zero model)         &               & $0.67^{**}$   &              \\
                              &               & $(0.28)$      &              \\
\hline
R$^2$                         & $0.03$        & $ $            & $ $           \\
Adj. R$^2$                    & $0.02$        & $ $            & $ $           \\
Num. obs.                     & $280$         & $290$         & $4566$       \\
AIC                           & $ $            & $2194.81$     & $5579.27$    \\
Log Likelihood                & $ $            & $-1091.41$    & $-2784.64$   \\
BIC                           & $ $            & $ $            & $5611.40$    \\
Num. groups: organisation     & $ $            & $ $            & $199$        \\
Num. groups: item             & $ $            & $ $            & $41$         \\
Var: organisation (Intercept) & $ $            & $ $            & $0.32$       \\
Var: item (Intercept)         & $ $            & $ $            & $0.65$       \\
\hline
\end{tabular}
\begin{tablenotes}[flushleft]
\scriptsize{\item[\hspace{-5mm}] $^{***}p<0.01$; $^{**}p<0.05$; $^{*}p<0.1$.
\item[\hspace{-5mm}] \textit{Note}: Instead of \texttt{Political budget} as the main explanatory variable, the table features the results for the entire budget of an organization. \textit{Self-perceived influence} is estimated using OLS; the model for \textit{reputational influence} uses maximum likelihood with a poisson regression for the count and a binomial regression with logit link for the zero model; \textit{preference attainment} is estimated using a mixed model fit by maximum likelihood with random intercepts for organizations and items.}
\end{tablenotes}
\end{threeparttable}
\caption{Regression results - comparison of influence measures using the overall budget to measure financial resources. }
\label{app_t_reg.compare.budget} 
\end{table}

\begin{table}[!h]\centering\footnotesize
\begin{threeparttable}
\renewcommand{\arraystretch}{1.2}
\begin{tabular}{l c c}
 & Preference attainment (1) & Preference attainment (2)\\
\hline
\noalign{\vskip 0.1cm}
Intercept                     & $0.66^{***}$ & $0.83^{***}$ \\
                              & $(0.10)$     & $(0.17)$     \\
Political budget              & $-0.04^{*}$  & $-0.06^{**}$ \\
                              & $(0.02)$     & $(0.03)$     \\
Business interests                      & $-0.19^{**}$ & $-0.25^{**}$ \\
                              & $(0.09)$     & $(0.11)$     \\
\hline
AIC                           & $5582.54$    & $5070.42$    \\
BIC                           & $5607.91$    & $5102.03$    \\
Log Likelihood                & $-2787.27$   & $-2530.21$   \\
Num. obs.                     & $4200$       & $4112$       \\
Num. groups: organization     & $187$        & $163$        \\
Var: organization (Intercept) & $0.17$       & $0.25$       \\
Num. groups: item             & $ $           & $41$         \\
Var: item (Intercept)         & $ $           & $0.63$       \\
\hline
\end{tabular}
\begin{tablenotes}[flushleft]
\scriptsize{\item[\hspace{-5mm}] $^{***}p<0.01$; $^{**}p<0.05$; $^{*}p<0.1$.
\item[\hspace{-5mm}] \textit{Note}: \textit{Model 1} features the estimates for an alternative specification without the random intercept for items. \textit{Model 2} features the results for the subsample of organizations for which the \textit{preference attainment} measure builds on more than 10 items. \textit{Preference attainment} is estimated using a mixed model fit by maximum likelihood with random intercepts for organizations and - in \textit{Model 2} - items.}
\end{tablenotes}
\end{threeparttable}

\caption{\textit{Preference attainment} regression results - additional estimations with random intercept only for organization (\textit{Model 1}) and on a subsample of organizations with more than 10 observations (\textit{Model 2}).}
\label{app_t_reg.compare.pa} 
\end{table}

\begin{table}[!h]\centering\footnotesize
\begin{threeparttable}
\renewcommand{\arraystretch}{1.2}
\begin{tabular}{l c c}
& Preference attainment (1) & Preference attainment (2)\\
\hline
\noalign{\vskip 0.1cm}
Intercept                          & $1.12^{***}$ & $0.68^{***}$ \\
                                   & $(0.34)$     & $(0.17)$     \\
Political budget                   & $-0.07$      & $-0.04^{*}$  \\
                                   & $(0.05)$     & $(0.03)$     \\
Business interests                           & $-0.47^{**}$ & $-0.14$      \\
                                   & $(0.20)$     & $(0.11)$     \\
\hline
AIC                                & $1646.93$    & $3563.95$    \\
BIC                                & $1673.05$    & $3593.69$    \\
Log Likelihood                     & $-818.47$    & $-1776.98$   \\
Num. obs.                          & $1371$       & $2829$       \\
Num. groups: organization          & $91$         & $163$        \\
Var: organization (Intercept)      & $15$         & $26$         \\
Var: name.organisation (Intercept) & $0.49$       & $0.15$       \\
Var: item (Intercept)              & $0.85$       & $0.50$       \\
\hline
\end{tabular}
\begin{tablenotes}[flushleft]
\scriptsize{\item[\hspace{-5mm}] $^{***}p<0.01$; $^{**}p<0.05$; $^{*}p<0.1$.
\item[\hspace{-5mm}] \textit{Note}: \textit{Model 1} features the estimates for the subsample from the 'Klimastrategie nach 2020' consultation. \textit{Model 2} features the estimates for the consultation on 'Energiestrategie 2050'. \textit{Preference attainment} is estimated using a mixed model fit by maximum likelihood with random intercepts for organizations and items.}
\end{tablenotes}
\end{threeparttable}

\caption{\textit{Preference attainment} regression results - separate estimations for the 'Klimapolitik nach 2020' (\textit{Model 1}) and 'Energiestrategie 20250' (\textit{Model 2}).}
\label{app_t_reg.compare.pa_indiv} 
\end{table}


\clearpage
\section{Appendix C}\label{app_bootstrap}

\begin{figure}[hbt!]
\includegraphics[width=.915\textwidth]{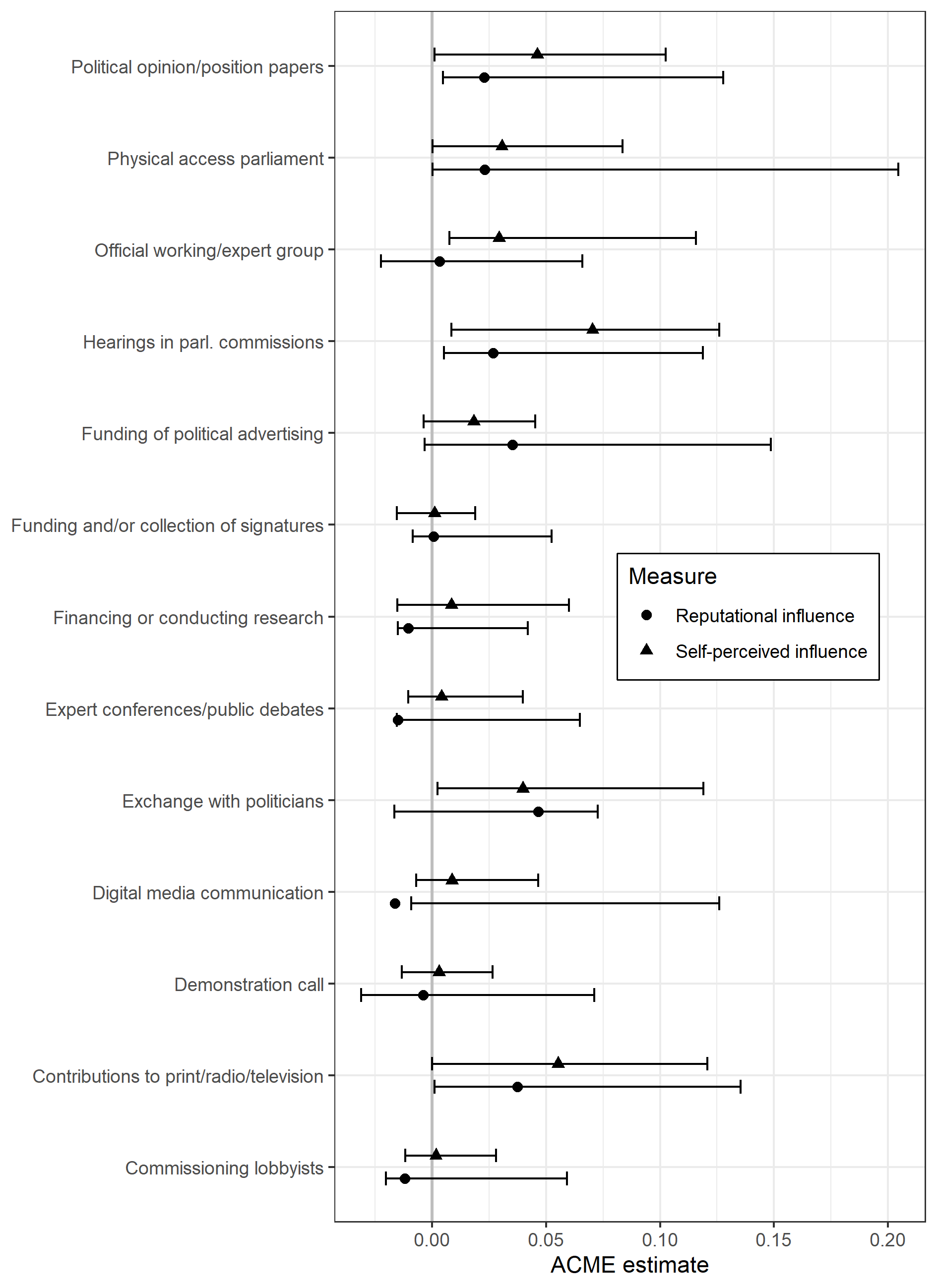}
\caption{Comparison of ACME estimates with bootstrapped 95\% confidence intervals. }
\label{app_f_coefplot.boot}
\floatnote{95\% confidence intervals estimated using nonparametric bootstrap. Estimates are not available for the \textit{preference attainment} measure.}
\end{figure}

\begin{table}[!h]\centering
\centering
\resizebox{0.96\textwidth}{!}{
\renewcommand{\arraystretch}{1.2}
\begin{tabular}[t]{>{\raggedright\arraybackslash}p{12em}cccccccccc}
\multicolumn{1}{l}{\em{ }} & \multicolumn{5}{l}{\em{Self-perceived}} & \multicolumn{5}{l}{\em{Reputational}} \\
\cmidrule(l{3pt}r{3pt}){2-6} \cmidrule(l{3pt}r{3pt}){7-11}
Mediator & ATE & ADE & ACME & CI 95\% ACME & Mediated & ATE & ADE & ACME & CI 95\% ACME & Mediated\\
\midrule
Exchange with politicians & 0.242 & 0.203 & 0.040 & {}[0.002, 0.119] & 0.164 & 0.369 & 0.322 & 0.047 & {}[-0.017, 0.073] & 0.126\\
Expert conferences/public debates & 0.225 & 0.221 & 0.004 & {}[-0.011, 0.040] & 0.018 & 0.304 & 0.319 & -0.015 & {}[-0.015, 0.065] & -0.049\\
Funding of political advertising & 0.214 & 0.196 & 0.018 & {}[-0.004, 0.045] & 0.085 & 0.282 & 0.247 & 0.035 & {}[-0.003, 0.149] & 0.125\\
Funding and/or collection of signatures & 0.230 & 0.228 & 0.001 & {}[-0.015, 0.019] & 0.005 & 0.317 & 0.317 & 0.001 & {}[-0.008, 0.052] & 0.002\\
Demonstration call & 0.229 & 0.226 & 0.003 & {}[-0.013, 0.027] & 0.013 & 0.305 & 0.309 & -0.004 & {}[-0.031, 0.071] & -0.013\\
Commissioning lobbyists & 0.229 & 0.227 & 0.002 & {}[-0.012, 0.028] & 0.007 & 0.309 & 0.321 & -0.012 & {}[-0.020, 0.059] & -0.039\\
Physical access parliament & 0.215 & 0.184 & 0.031 & {}[0.000, 0.084] & 0.142 & 0.265 & 0.242 & 0.023 & {}[0.000, 0.205] & 0.087\\
Hearings in parl. commissions & 0.220 & 0.149 & 0.070 & {}[0.008, 0.126] & 0.320 & 0.288 & 0.261 & 0.027 & {}[0.005, 0.119] & 0.093\\
Official working/expert group & 0.204 & 0.174 & 0.029 & {}[0.008, 0.116] & 0.145 & 0.317 & 0.314 & 0.003 & {}[-0.022, 0.066] & 0.010\\
Political opinion/position papers & 0.243 & 0.197 & 0.046 & {}[0.001, 0.102] & 0.190 & 0.315 & 0.292 & 0.023 & {}[0.005, 0.128] & 0.072\\
Financing or conducting research & 0.221 & 0.212 & 0.008 & {}[-0.015, 0.060] & 0.038 & 0.310 & 0.320 & -0.011 & {}[-0.015, 0.042] & -0.034\\
Contributions to print/radio/television & 0.239 & 0.184 & 0.055 & {}[0.000, 0.121] & 0.231 & 0.329 & 0.292 & 0.037 & {}[0.001, 0.135] & 0.114\\
Digital media communication & 0.224 & 0.215 & 0.009 & {}[-0.007, 0.047] & 0.039 & 0.274 & 0.290 & -0.016 & {}[-0.009, 0.126] & -0.060\\
\bottomrule
\multicolumn{11}{l}{\rule{0pt}{1em}\textit{Note: } 95\% confidence intervals estimated using nonparametric bootstrap. Estimates are not available for the \textit{preference attainment} measure.}\\
\end{tabular}}
\caption{Mediation analysis estimates with bootstrapped 95\% confidence intervals.}
\label{app_t_mediation.boot} 
\end{table}


\clearpage
\section{Appendix D}\label{app_sensitivity}

While the original mediation package includes sensitivity analyses for pretreatment confounding, unfortunately it is restricted to linear and binary probit models \citep{Tingley2014Mediation:Analysis}. However, we present results for the sensitivity analyses of potential posttreatment confounding.

In contrast to pretreatment confounding, which can be addressed by including relevant covariates, the sequential ignorability assumption technically prohibits any posttreatment confounding, irrespective of whether such a variable is observed or not.

Provided the two mediators are causally independent, there is no advantage to consider them simultaneously as opposed to a one-by-one mediation analysis \citep{Imai2013IdentificationExperiments}. Hence, as a first step of our sensitivity analysis, we test whether the mediators, i.e., the lobbying activities, are conditionally independent. We restrict the set of potential posttreatment confounders to those activities, which significantly mediate the effect of \textit{political budget} on influence themselves.

\begin{table}[!h]\centering\footnotesize
\begin{threeparttable}
\renewcommand{\arraystretch}{1.2}
\begin{tabular}{l c c c c}
 & \multicolumn{4}{c}{Physical access parliament} \\
\cmidrule(lr){2-5}
 & Model 1 & Model 2 & Model 3 & Model 4 \\
\midrule
(Intercept)                             & $-1.36^{***}$ & $-1.39^{***}$ & $-1.85^{***}$ & $-1.92^{***}$ \\
                                        & $(0.19)$      & $(0.19)$      & $(0.24)$      & $(0.24)$      \\
Hearings in parl. commissions           & $1.17^{***}$  &               &               &               \\
                                        & $(0.20)$      &               &               &               \\
Official working/expert group           &               & $0.35^{*}$    &               &               \\
                                        &               & $(0.18)$      &               &               \\
Political opinion/position papers       &               &               & $1.01^{***}$  &               \\
                                        &               &               & $(0.22)$      &               \\
Contributions to print/radio/television &               &               &               & $1.06^{***}$  \\
                                        &               &               &               & $(0.20)$      \\
Political budget                        & $0.17^{***}$  & $0.25^{***}$  & $0.21^{***}$  & $0.22^{***}$  \\
                                        & $(0.05)$      & $(0.05)$      & $(0.05)$      & $(0.05)$      \\
Business interests                      & $-0.17$       & $-0.06$       & $-0.01$       & $0.14$        \\
                                        & $(0.18)$      & $(0.18)$      & $(0.18)$      & $(0.18)$      \\
\midrule
AIC                                     & $256.93$      & $288.20$      & $267.65$      & $262.50$      \\
BIC                                     & $271.31$      & $302.58$      & $282.03$      & $276.88$      \\
Log Likelihood                          & $-124.47$     & $-140.10$     & $-129.83$     & $-127.25$     \\
Deviance                                & $248.93$      & $280.20$      & $259.65$      & $254.50$      \\
Num. obs.                               & $269$         & $269$         & $269$         & $269$         \\
\hline
\end{tabular}
\begin{tablenotes}[flushleft]
\scriptsize{\item[\hspace{-5mm}] $^{***}p<0.01$; $^{**}p<0.05$; $^{*}p<0.1$.}
\end{tablenotes}
\end{threeparttable}
\caption{Conditional independence of the lobbying activity "Physical access to parliament".}
\label{app_t_condind.polact3} 
\end{table}

\begin{table}[!h]\centering\footnotesize
\begin{threeparttable}
\renewcommand{\arraystretch}{1.2}
\begin{tabular}{l c c c c}
 & \multicolumn{4}{c}{Hearings in parl. commissions} \\
\cmidrule(lr){2-5}
 & Model 1 & Model 2 & Model 3 & Model 4 \\
\midrule
(Intercept)                             & $-1.94^{***}$ & $-1.93^{***}$ & $-2.11^{***}$ & $-2.10^{***}$ \\
                                        & $(0.22)$      & $(0.22)$      & $(0.25)$      & $(0.24)$      \\
Physical access parliament              & $1.16^{***}$  &               &               &               \\
                                        & $(0.20)$      &               &               &               \\
Official working/expert group           &               & $0.89^{***}$  &               &               \\
                                        &               & $(0.19)$      &               &               \\
Political opinion/position papers       &               &               & $0.76^{***}$  &               \\
                                        &               &               & $(0.21)$      &               \\
Contributions to print/radio/television &               &               &               & $0.74^{***}$  \\
                                        &               &               &               & $(0.20)$      \\
Political budget                        & $0.26^{***}$  & $0.27^{***}$  & $0.29^{***}$  & $0.29^{***}$  \\
                                        & $(0.05)$      & $(0.05)$      & $(0.05)$      & $(0.05)$      \\
Business interests                      & $0.35^{*}$    & $0.26$        & $0.33^{*}$    & $0.42^{**}$   \\
                                        & $(0.19)$      & $(0.19)$      & $(0.19)$      & $(0.19)$      \\
\midrule
AIC                                     & $238.61$      & $250.89$      & $259.72$      & $259.12$      \\
BIC                                     & $252.98$      & $265.27$      & $274.10$      & $273.50$      \\
Log Likelihood                          & $-115.30$     & $-121.45$     & $-125.86$     & $-125.56$     \\
Deviance                                & $230.61$      & $242.89$      & $251.72$      & $251.12$      \\
Num. obs.                               & $269$         & $269$         & $269$         & $269$         \\
\hline
\end{tabular}
\begin{tablenotes}[flushleft]
\scriptsize{\item[\hspace{-5mm}] $^{***}p<0.01$; $^{**}p<0.05$; $^{*}p<0.1$.}
\end{tablenotes}
\end{threeparttable}
\caption{Conditional independence of the lobbying activity "Hearings in parl. commissions".}
\label{app_t_condind.polact4} 
\end{table}

\begin{table}[!h]\centering\footnotesize
\begin{threeparttable}
\renewcommand{\arraystretch}{1.2}
\begin{tabular}{l c c c c}
 & \multicolumn{4}{c}{Official working/expert group} \\
\cmidrule(lr){2-5}
 & Model 1 & Model 2 & Model 3 & Model 4 \\
\midrule
(Intercept)                             & $-1.07^{***}$ & $-1.04^{***}$ & $-1.28^{***}$ & $-1.41^{***}$ \\
                                        & $(0.18)$      & $(0.18)$      & $(0.20)$      & $(0.20)$      \\
Physical access parliament              & $0.34^{*}$    &               &               &               \\
                                        & $(0.19)$      &               &               &               \\
Hearings in parl. commissions           &               & $0.94^{***}$  &               &               \\
                                        &               & $(0.20)$      &               &               \\
Political opinion/position papers       &               &               & $0.54^{***}$  &               \\
                                        &               &               & $(0.18)$      &               \\
Contributions to print/radio/television &               &               &               & $0.75^{***}$  \\
                                        &               &               &               & $(0.18)$      \\
Political budget                        & $0.25^{***}$  & $0.19^{***}$  & $0.24^{***}$  & $0.23^{***}$  \\
                                        & $(0.05)$      & $(0.05)$      & $(0.05)$      & $(0.05)$      \\
Business interests                      & $0.14$        & $0.05$        & $0.16$        & $0.26$        \\
                                        & $(0.17)$      & $(0.17)$      & $(0.17)$      & $(0.17)$      \\
\midrule
AIC                                     & $325.47$      & $305.77$      & $319.77$      & $310.44$      \\
BIC                                     & $339.85$      & $320.15$      & $334.15$      & $324.82$      \\
Log Likelihood                          & $-158.73$     & $-148.89$     & $-155.88$     & $-151.22$     \\
Deviance                                & $317.47$      & $297.77$      & $311.77$      & $302.44$      \\
Num. obs.                               & $269$         & $269$         & $269$         & $269$         \\
\hline
\end{tabular}
\begin{tablenotes}[flushleft]
\scriptsize{\item[\hspace{-5mm}] $^{***}p<0.01$; $^{**}p<0.05$; $^{*}p<0.1$.}
\end{tablenotes}
\end{threeparttable}
\caption{Conditional independence of the lobbying activity "Official working/expert group".}
\label{app_t_condind.polact5} 
\end{table}

\begin{table}[!h]\centering\footnotesize
\begin{threeparttable}
\renewcommand{\arraystretch}{1.2}
\begin{tabular}{l c c c c}
 & \multicolumn{4}{c}{Political opinion/position papers} \\
\cmidrule(lr){2-5}
 & Model 1 & Model 2 & Model 3 & Model 4 \\
\midrule
(Intercept)                             & $-0.47^{***}$ & $-0.34^{**}$ & $-0.45^{***}$ & $-0.79^{***}$ \\
                                        & $(0.18)$      & $(0.17)$     & $(0.17)$      & $(0.20)$      \\
Physical access parliament              & $1.11^{***}$  &              &               &               \\
                                        & $(0.23)$      &              &               &               \\
Hearings in parl. commissions           &               & $0.77^{***}$ &               &               \\
                                        &               & $(0.22)$     &               &               \\
Official working/expert group           &               &              & $0.55^{***}$  &               \\
                                        &               &              & $(0.18)$      &               \\
Contributions to print/radio/television &               &              &               & $0.87^{***}$  \\
                                        &               &              &               & $(0.18)$      \\
Political budget                        & $0.23^{***}$  & $0.22^{***}$ & $0.25^{***}$  & $0.24^{***}$  \\
                                        & $(0.05)$      & $(0.05)$     & $(0.05)$      & $(0.05)$      \\
Business interests                      & $-0.16$       & $-0.24$      & $-0.20$       & $-0.00$       \\
                                        & $(0.17)$      & $(0.17)$     & $(0.17)$      & $(0.18)$      \\
\midrule
AIC                                     & $297.24$      & $310.51$     & $314.04$      & $298.41$      \\
BIC                                     & $311.62$      & $324.88$     & $328.42$      & $312.79$      \\
Log Likelihood                          & $-144.62$     & $-151.25$    & $-153.02$     & $-145.20$     \\
Deviance                                & $289.24$      & $302.51$     & $306.04$      & $290.41$      \\
Num. obs.                               & $269$         & $269$        & $269$         & $269$         \\
\hline
\end{tabular}
\begin{tablenotes}[flushleft]
\scriptsize{\item[\hspace{-5mm}] $^{***}p<0.01$; $^{**}p<0.05$; $^{*}p<0.1$.}
\end{tablenotes}
\end{threeparttable}
\caption{Conditional independence of the lobbying activity "Political opinion/position papers".}
\label{app_t_condind.polact6} 
\end{table}

\begin{table}[!h]\centering\footnotesize
\begin{threeparttable}
\renewcommand{\arraystretch}{1.2}
\begin{tabular}{l c c c c}
 & \multicolumn{4}{c}{Contributions to print/radio/television} \\
\cmidrule(lr){2-5}
 & Model 1 & Model 2 & Model 3 & Model 4 \\
\midrule
(Intercept)                       & $-0.26$       & $-0.17$       & $-0.31^{*}$   & $-0.54^{***}$ \\
                                  & $(0.17)$      & $(0.17)$      & $(0.17)$      & $(0.18)$      \\
Physical access parliament        & $1.13^{***}$  &               &               &               \\
                                  & $(0.22)$      &               &               &               \\
Hearings in parl. commissions     &               & $0.79^{***}$  &               &               \\
                                  &               & $(0.21)$      &               &               \\
Official working/expert group     &               &               & $0.76^{***}$  &               \\
                                  &               &               & $(0.18)$      &               \\
Political opinion/position papers &               &               &               & $0.86^{***}$  \\
                                  &               &               &               & $(0.18)$      \\
Political budget                  & $0.17^{***}$  & $0.17^{***}$  & $0.18^{***}$  & $0.17^{***}$  \\
                                  & $(0.05)$      & $(0.05)$      & $(0.05)$      & $(0.05)$      \\
Business interests                & $-0.57^{***}$ & $-0.60^{***}$ & $-0.58^{***}$ & $-0.51^{***}$ \\
                                  & $(0.17)$      & $(0.17)$      & $(0.17)$      & $(0.17)$      \\
\midrule
AIC                               & $304.65$      & $320.50$      & $316.69$      & $311.26$      \\
BIC                               & $319.03$      & $334.87$      & $331.07$      & $325.64$      \\
Log Likelihood                    & $-148.33$     & $-156.25$     & $-154.35$     & $-151.63$     \\
Deviance                          & $296.65$      & $312.50$      & $308.69$      & $303.26$      \\
Num. obs.                         & $269$         & $269$         & $269$         & $269$         \\
\hline
\end{tabular}
\begin{tablenotes}[flushleft]
\scriptsize{\item[\hspace{-5mm}] $^{***}p<0.01$; $^{**}p<0.05$; $^{*}p<0.1$.}
\end{tablenotes}
\end{threeparttable}
\caption{Conditional independence of the lobbying activity "Contributions to print/radio/television".}
\label{app_t_condind.polact8} 
\end{table}

\begin{table}[!h]\centering\footnotesize
\begin{threeparttable}
\renewcommand{\arraystretch}{1.2}
\begin{tabular}{@{}l@{} c c c c c}
 & \multicolumn{1}{c}{Physical access} & \multicolumn{1}{c}{Hearings in parl.} & \multicolumn{1}{c}{Official working /} & \multicolumn{1}{c}{Political opinion /} & \multicolumn{1}{c}{Contributions to}\\
 & parliament & commissions & expert group & position papers & print/radio/television\\

\midrule
(Intercept)        & $-1.33^{***}$ & $-1.70^{***}$ & $-1.04^{***}$ & $-0.37^{**}$ & $-0.20$       \\
                   & $(0.19)$      & $(0.20)$      & $(0.18)$      & $(0.17)$     & $(0.17)$      \\
Political budget   & $0.28^{***}$  & $0.34^{***}$  & $0.28^{***}$  & $0.29^{***}$ & $0.25^{***}$  \\
                   & $(0.04)$      & $(0.04)$      & $(0.04)$      & $(0.05)$     & $(0.04)$      \\
Business interests & $-0.04$       & $0.29$        & $0.13$        & $-0.17$      & $-0.53^{***}$ \\
                   & $(0.17)$      & $(0.18)$      & $(0.17)$      & $(0.17)$     & $(0.17)$      \\
\midrule
AIC                & $289.79$      & $271.51$      & $326.77$      & $321.48$     & $333.20$      \\
BIC                & $300.58$      & $282.29$      & $337.56$      & $332.27$     & $343.98$      \\
Log Likelihood     & $-141.90$     & $-132.75$     & $-160.39$     & $-157.74$    & $-163.60$     \\
Deviance           & $283.79$      & $265.51$      & $320.77$      & $315.48$     & $327.20$      \\
Num. obs.          & $269$         & $269$         & $269$         & $269$        & $269$         \\
\hline
\end{tabular}
\begin{tablenotes}[flushleft]
\scriptsize{\item[\hspace{-5mm}] $^{***}p<0.01$; $^{**}p<0.05$; $^{*}p<0.1$.}
\end{tablenotes}
\end{threeparttable}
\caption{Conditional independence test of the lobbying activities.}
\label{app_t_condind.all} 
\end{table}

The results from Table \ref{app_t_condind.polact3} to \ref{app_t_condind.polact8} in combination with Table \ref{app_t_condind.all} suggest that we cannot assume conditional independence for any of the five mediators considered. In such a case of multiple causally dependent mediators, the identification of ACME, as outlined by \cite{Imai2013IdentificationExperiments}, requires an additional assumption: To ensure unbiased estimates, any interaction between \textit{political budget} and the mediator of interest must be ruled out. Since this must hold for every observation, the assumption is unlikely to hold. Rather, \cite{Imai2013IdentificationExperiments} suggest considering two sensitivity parameters, which allow to relax the assumption (one representing the direction of the interaction effect and the other the degree of heterogeneity in the treatment–mediator interaction).

Unfortunately, again, this type of analysis is available for a limited set of models only. Hence, we refer to the sensitivity analysis proposed by \cite{Cheng2018MediationData} to further analyze potential posttreatment confounding. Their approach includes the confounder as an independent variable in the mediator and outcome model. Further, it requires the estimation of an additional model to estimate the confounder itself using \textit{political budget} and \textit{business interests} as predictors. The coefficient for \textit{political budget} from this confounder model serves as a reference value for the sensitivity analysis, as the corresponding graphs illustrate how the ACME varies depending on values within a range of this reference value.

\begin{figure}[hbt!]
\includegraphics[width=.935\textwidth]{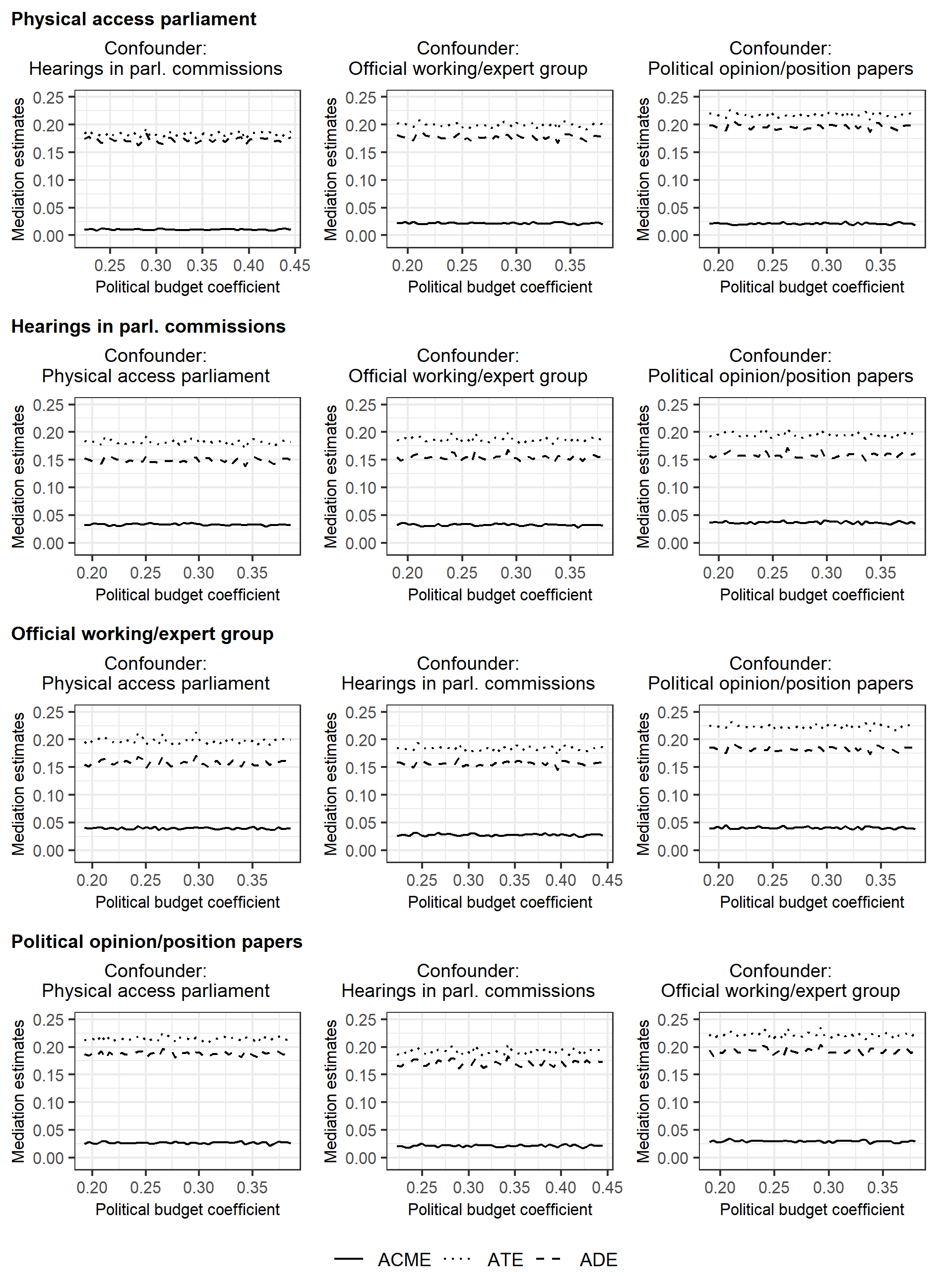}
\caption{Sensitivity analysis for mediators of political budget on self-perceived influence.}
\label{app_f_sens_self}
\floatnote{ACME stands for the mediated effect of political budget, ADE for the direct and ATE for the total effect.}
\end{figure}

\begin{figure}[hbt!]
\includegraphics[width=.935\textwidth]{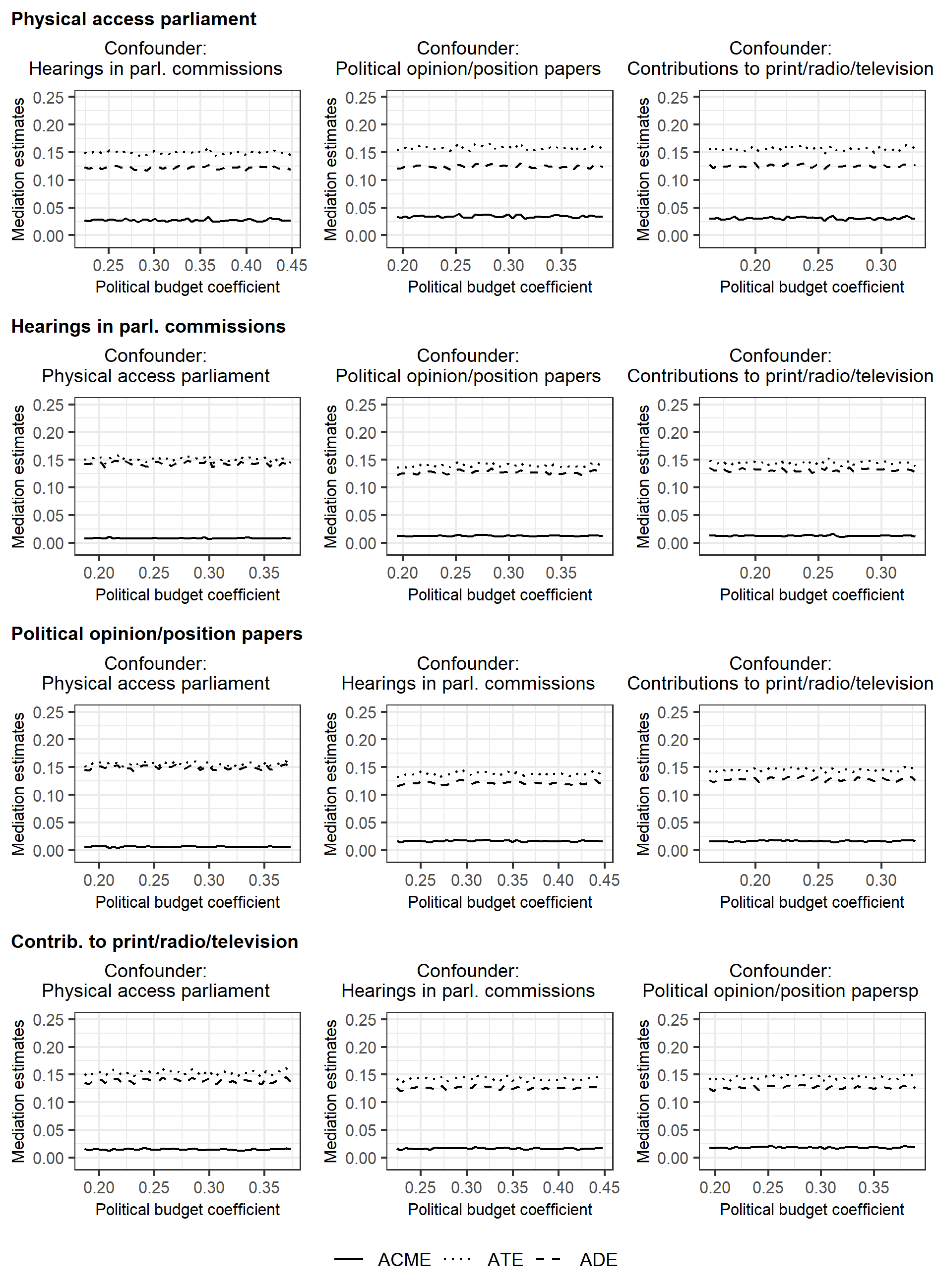}
\caption{Sensitivity analysis for mediators of political budget on reputational influence.}
\label{app_f_sens_rep}
\floatnote{ACME stands for the mediated effect of political budget, ADE for the direct and ATE for the total effect.}
\end{figure}

The graphs in Figure \ref{app_f_sens_self} and \ref{app_f_sens_rep} generally do not reveal a trend in the ACMEs along the x-axis, which depicts the variation of the \textit{political budget} estimate. There is no golden rule on the degree of violation that would delegitimize our findings. However, when interpreting the results from the mediation analyses, it is important to keep in mind the strictness of the identification assumption. It follows, that conclusions need to be drawn with due caution.

\clearpage

\end{document}